\PassOptionsToPackage{hyperfootnotes=false}{hyperref}
\documentclass[a4paper,fleqn]{cas-sc}

\usepackage{booktabs}
\usepackage{rotating}
\usepackage{longtable}
\usepackage[numbers,sort&compress]{natbib}
\usepackage{amsmath,amssymb,amsfonts}
\usepackage{bm}
\usepackage{physics}
\usepackage{graphicx}
\usepackage{subcaption}
\usepackage{float}
\usepackage{placeins}
\usepackage{booktabs}
\usepackage{multirow}
\usepackage{array}
\usepackage{cleveref}
\usepackage{enumitem}
\usepackage{xcolor}
\usepackage{algorithm}
\usepackage{algpseudocode}
\usepackage{listings}

\newcommand{\tabnote}[1]{%
  \par\smallskip
  \parbox{0.96\linewidth}{\footnotesize\raggedright\textit{Note:} #1}%
}

\newdimen\casSavedHfuzz

\hypersetup{
  colorlinks=true,
  linkcolor=blue,
  citecolor=blue,
  urlcolor=blue
}

\begin{document}
\let\WriteBookmarks\relax
\renewcommand\floatpagefraction{.6}
\renewcommand\textfraction{.05}

\shorttitle{}
\shortauthors{Spyrou et al.}

\title[mode=title]{Probabilistic Physics-Informed Neural Solvers for Woods–Saxon Parameter Identification: A Coupled Forward–Inverse Approach}

\author[1]{Iraklis Spyrou}[orcid=0009-0000-1347-5345]
\cormark[1]
\ead{i.spyrou@iit.demokritos.gr}
\credit{Conceptualization, Methodology, Software, Validation, Formal analysis, Investigation, Data curation, Visualization, Writing -- original draft, Writing -- review and editing}

\author[3]{Christos Tsepas}
\credit{Conceptualization, Methodology, Software, Validation, Formal analysis, Investigation, Writing -- review and editing}

\author[2]{Vaia Prassa}
\fnmark[1]
\credit{Supervision, Conceptualization, Writing -- review and editing}

\author[1]{Christoforos Rekatsinas}
\fnmark[1]
\credit{Supervision, Conceptualization, Project administration, Writing -- review and editing}

\affiliation[1]{organization={INSANE Group, IIT, NCSR Demokritos},
  city={Athens},
  country={Greece}}

\affiliation[2]{organization={Department of Physics, University of Thessaly},
  city={Lamia},
  country={Greece}}

\affiliation[3]{organization={Independent Researcher},
  city={Utrecht},
  country={The Netherlands}}

\cortext[1]{Corresponding author}
\fntext[1]{These authors contributed equally.}

\begin{abstract}
The Woods--Saxon mean-field approach provides a compact description of bound single-particle motion in finite nuclei, while physics-informed neural networks offer a differentiable route for solving the associated inverse quantum problem from sparse spectral data. This study develops a probabilistic physics-informed framework in which WaveNet represents the separated single-particle wavefunction and ParamNet maps selected spectra, nuclear descriptors, and quantum numbers to a learned output distribution over six global Woods--Saxon parameters. The Hamiltonian includes the central Woods--Saxon interaction, the proton Coulomb term, and the spin--orbit contribution, while the training process enforces spectral energy consistency, Schrödinger-equation residuals, boundary conditions, normalization, orthogonality, spin--orbit splitting constraints, and latent regularization; the distribution-mean estimator serves as the primary selection-free parameter estimate and is validated with an independent finite-difference radial solver. Synthetic closure tests based on the Seminole and Wahlborn parameterizations recover all six parameters with sub-percent relative errors and reproduce the reference spectra with mean absolute deviations of \(0.0109~\mathrm{MeV}\) and \(0.0131~\mathrm{MeV}\), respectively. For experimental spectra using the Wahlborn potential expression, the distribution-mean estimator reduces the all-state mean absolute error from \(1.0783\) to \(0.8303~\mathrm{MeV}\). Using the Seminole expression, the distribution-mean estimator attains an all-state mean absolute error of \(0.8068~\mathrm{MeV}\), close to the \(0.7969~\mathrm{MeV}\) obtained with the Seminole reference parameters. This comparable accuracy is achieved using \(42\) selected experimental levels, approximately \(51\%\) fewer spectral targets than those used in the Seminole reference calibration. The results indicate that sparse but structured single-particle spectra can constrain global Woods--Saxon interactions within a complementary differentiable framework that addresses both the forward eigenvalue problem and inverse parameter identification while supplying a learned output spread as a model-derived, qualitative measure of parameter stiffness.
\end{abstract}


\casSavedHfuzz=\hfuzz
\hfuzz=.25\textwidth
\maketitle
\hfuzz=\casSavedHfuzz

\section{Introduction}

The time-independent Schrödinger equation in three spatial dimensions
connects the spatial structure of a quantum state, the interaction
potential, and the corresponding energy spectrum. In nuclear physics,
this connection is central to effective descriptions of single-nucleon
motion in finite nuclei. Although the complete nuclear many-body
problem is considerably more complex, a widely used approximation
represents the interaction of one nucleon with the remaining nucleons
through an average mean-field potential.

Among phenomenological mean-field models, the Woods--Saxon potential is
widely used in nuclear-structure and reaction calculations. Its finite
depth and diffuse surface provide a compact representation of the
finite nuclear density profile, particularly where surface effects are
important \cite{schwierz2007parameterization}. In single-particle
applications, the central interaction is supplemented by a Coulomb term
for protons and a spin--orbit term. The Coulomb contribution describes
the electrostatic interaction between a proton and the nuclear core,
whereas the spin--orbit interaction is required to reproduce the
observed splitting of states with the same orbital angular momentum and
different total angular momentum \cite{nedjadi1988relativistic}.

Determining Woods--Saxon parameters from nuclear observables is a
classical inverse problem. Given single-particle and single-hole
energies with assigned quantum numbers, one seeks the central depth,
isospin dependence, radius, diffuseness, and spin--orbit parameters that
reproduce the observed spectrum. The Seminole parameterization, for
example, was obtained through a global least-squares calibration to
approximately \(86\) adopted orbital energies around seven doubly magic
nuclei spanning \(^{16}\mathrm{O}\) to \(^{208}\mathrm{Pb}\)
\cite{schwierz2007parameterization}. Such deterministic fits are
physically interpretable and computationally efficient when the
potential form, state assignments, and calibration data are fixed, but
they primarily return a single optimum. The inverse map may also be
ill-conditioned: for Woods--Saxon-type wells, a reduction in radius can
be offset by an increase in potential depth while approximately
preserving average single-particle binding energies
\cite{kahane1989testing}. Sparse, noisy, or incomplete level schemes can
therefore leave compensating parameter directions weakly constrained,
making the characterization of admissible parameter variability an
important part of the inverse problem.

Physics-informed neural networks (PINNs) provide an alternative route
for forward and inverse problems governed by differential equations.
They incorporate governing equations and physical constraints directly
into the training objective, reducing the dependence on dense labeled
solution data \cite{raissi2019physics}. For quantum
eigenvalue problems, PINN formulations have been developed to learn
eigenfunctions and eigenvalues while enforcing the Schrödinger-equation
residual, boundary conditions, normalization, and orthogonality
\cite{jin2022physics}. These additional constraints are essential for
distinguishing physically admissible quantum states rather than merely
reducing a differential-equation residual. Most related quantum
applications have emphasized forward problems with prescribed
potentials. Energy-conditioned solvers have recovered ground and
excited states for several one-dimensional systems, including the
Woods--Saxon potential \cite{wu2025energy}, while inverse PINNs have
identified unknown coefficients in Schrödinger-type equations
\cite{li2021scarf}. These studies do not directly address the global
recovery of a nuclear single-particle Woods--Saxon interaction from a
reduced set of bound-state spectra.

A further limitation of deterministic inverse methods is that parameter
ambiguity is not represented directly by a single best-fit vector.
Bayesian PINNs address uncertainty by combining physical constraints
with posterior inference over neural-network weights or unknown model
quantities \cite{yang2020bpinn}. Such weight-space inference can require
costly sampling or variational approximations in a high-dimensional
parameter space. Output-space variational methods provide a more
efficient alternative by learning distributions directly for the
quantities of interest rather than for every network weight
\cite{wei2024vifo}. This distinction is particularly relevant for the
present inverse problem, where the quantities of interest are the
physically interpretable Woods--Saxon parameters. The resulting output
spread can indicate model-derived confidence and qualitative parameter
stiffness, although it should not be interpreted as a calibrated
Bayesian credible interval without additional calibration tests.

This work develops a probabilistic physics-informed framework to address
both the forward single-particle eigenvalue problem and the inverse
identification of a global Woods--Saxon interaction from bound-state
spectra. Spectral observations and physical constraints are combined so
that the framework recovers physically consistent wavefunctions and
energies while producing a learned output distribution over the six
interaction parameters. The distribution mean is used as the primary and the associated spread provides a
qualitative measure of how strongly the supplied spectra and physical
losses constrain each parameter. Architectural definitions, loss terms,
and implementation details are presented in the following sections.

The objective of the proposed framework is not to replace standard finite-difference solvers
for the forward radial Schrödinger equation. Rather, the aim is to construct a differentiable and probabilistic inverse solver in which spectral observations, physical constraints and parameter uncertainty are treated within a single trainable model. The method is evaluated using synthetic single-particle spectra generated from known Woods–Saxon parameterizations, allowing direct assessment of parameter recovery, uncertainty estimates and physical consistency. In addition, the inferred parameters are reintroduced into an independent finite-difference Hamiltonian as a
closure test, and the resulting spectrum is compared with the reference energies.

The central question addressed in this study is whether limited bound-state spectral information is sufficient to recover the underlying Woods–Saxon mean–field parameters while
maintaining physically consistent wavefunction representations and meaningful uncertainty estimates. By combining physics-informed learning with probabilistic parameter inference, the
proposed framework provides a computational route toward uncertainty-aware inverse modeling
of nuclear single-particle mean fields.

\section{Physical Background and Problem Formulation}
\label{sec:physical_background}

\subsection{Nuclear Mean-Field and Inverse Problem}
\label{subsec:mean_field_description}

Within the nuclear mean-field approximation, each nucleon moves in an
effective one-body potential generated by the remaining nucleons. For a
nucleus with mass number \(A\), proton number \(Z\), neutron number
\(N=A-Z\), and nucleon species \(\tau\in\{n,p\}\), the bound
single-particle states satisfy

\begin{equation}
\widehat{H}_{\tau}(\bm{\vartheta})
\Psi_k^{(\tau)}
=
E_k^{(\tau)}
\Psi_k^{(\tau)},
\label{eq:sp_schrodinger}
\end{equation}

where \(k\) denotes the quantum state and
\(\bm{\vartheta}\) contains the parameters of the effective
interaction. The inverse problem considered in this work is to infer one
global parameter distribution from a collection of experimental or
synthetic bound-state energies and their quantum-number assignments.

\subsection{Woods--Saxon Hamiltonian}
\label{subsec:woods_saxon_hamiltonian}

The single-particle Hamiltonian used throughout the forward solver and
the physics-informed losses is

\begin{equation}
\begin{aligned}
\widehat{H}_{\tau}(\bm{\vartheta})
={}&
-\frac{\hbar^{2}}{2\mu_{\tau}}\nabla^{2}
+
V_{\mathrm{N}}^{(\tau)}(r)
+
\delta_{\tau\mathrm{p}}V_{\mathrm{C}}(r)
&+
V_{\mathrm{SO}}^{(\tau)}(r)
\widehat{\bm{l}}\cdot\widehat{\bm{s}},
\end{aligned}
\label{eq:ws_hamiltonian}
\end{equation}

where \(\mu_{\tau}\) is the nucleon--core reduced mass and
\(\delta_{\tau\mathrm{p}}\) restricts the Coulomb contribution to
protons.

The nuclear central potential is represented by the Woods--Saxon form

\begin{equation}
V_{\mathrm{N}}^{(\tau)}(r)
=
-D_{\tau}(A,Z)
f(r;R,a),
\qquad
f(r;R,a)
=
\frac{1}
{1+\exp\left(\dfrac{r-R}{a}\right)},
\qquad
R=r_0A^{1/3}.
\label{eq:nuclear_ws_potential}
\end{equation}

Here, \(D_{\tau}\) is the isospin-dependent potential depth,
\(r_0\) is the central radius coefficient, and \(a\) is the surface
diffuseness.

Two expressions for the central depth are investigated. In the classical Wahlborn parameterization,

\begin{equation}
D_{\tau}^{\mathrm{W}}(A,Z)
=
V_0
\left[
1+
\eta_{\tau}\kappa
\frac{N-Z}{A}
\right],
\qquad
\eta_{\mathrm{p}}=+1,
\qquad
\eta_{\mathrm{n}}=-1,
\label{eq:wahlborn_central_depth}
\end{equation}

whereas the Seminole expression is

\begin{equation}
D_{\tau}^{\mathrm{S}}(A,Z)
=
V_0
\left[
1-
\frac{4\kappa}{A}
\left\langle
\bm{t}\cdot\bm{T}'
\right\rangle_{\tau}
\right].
\label{eq:seminole_central_depth}
\end{equation}

The isospin expectation value in
Eq.~\eqref{eq:seminole_central_depth} is evaluated according to 
\cite{schwierz2007parameterization}. Further details of the Wahlborn
and related Woods--Saxon parameterizations can be found in
Refs.~\cite{rost1968proton,dudek1979parameters,dudek1982description}.

For protons, the Coulomb field is approximated by that of a uniformly
charged spherical core,

\begin{equation}
V_{\mathrm{C}}(r)
=
\begin{cases}
\displaystyle
\frac{Z_{\mathrm{c}}e^2}{2R_{\mathrm{C}}}
\left[
3-
\left(
\frac{r}{R_{\mathrm{C}}}
\right)^2
\right],
&
r\leq R_{\mathrm{C}},
\\[3mm]
\displaystyle
\frac{Z_{\mathrm{c}}e^2}{r},
&
r>R_{\mathrm{C}},
\end{cases}
\label{eq:coulomb_potential_background}
\end{equation}

with \(Z_{\mathrm{c}}=Z-1\) and \(R_{\mathrm{C}}=R\).

The radial spin--orbit coefficient is

\begin{equation}
V_{\mathrm{SO}}^{(\tau)}(r)
=
\frac{(\hbar c)^2}
{2\mu_{\tau}^{2}r}
\frac{d}{dr}
\left[
\widetilde{D}_{\tau}
f(r;R_{\mathrm{SO}},a_{\mathrm{SO}})
\right],
\qquad
R_{\mathrm{SO}}
=
r_{0,\mathrm{SO}}A^{1/3},
\label{eq:spin_orbit_potential_background}
\end{equation}

where

\begin{equation}
\widetilde{D}_{\tau}
=
\begin{cases}
\lambda_{\mathrm{SO}}D_{\tau}^{\mathrm{W}}
& \text{for Wahlborn},\\[1mm]
\lambda_{\mathrm{SO}}V_0
& \text{for Seminole}.
\end{cases}
\label{eq:spin_orbit_depth}
\end{equation}

The spin--orbit diffuseness is set equal to the central diffuseness,
\(a_{\mathrm{SO}}=a\).

\subsection{Spherical Symmetry and Variable Separation}
\label{subsec:separated_formulation}

The Hamiltonian is assumed to be spherically symmetric, as in the
standard Woods--Saxon single-particle formulation
\cite{schwierz2007parameterization,junker2025susy}. After projection
onto a fixed \((l,j)\) channel, the spin--orbit operator is replaced by

\begin{equation}
\xi_{lj}
=
\frac{1}{2}
\left[
j(j+1)
-
l(l+1)
-
\frac{3}{4}
\right].
\label{eq:spin_orbit_eigenvalue_background}
\end{equation}

The spatial wavefunction is represented in separated form as

\begin{equation}
\Psi_{n_r l j m_l}^{(\tau)}
(r,\theta,\phi)
=
R_{n_r l j}^{(\tau)}(r)
\Theta_{l m_l}(\theta)
\Phi_{m_l}(\phi).
\label{eq:spatial_separable_representation}
\end{equation}

This representation gives the three differential equations whose
residuals are enforced during PINN training.

The radial equation is

\begin{equation}
\begin{aligned}
\Bigg[
&
-\frac{\hbar^2}{2\mu_{\tau}}
\left(
\frac{1}{r^2}
\frac{d}{dr}
r^2
\frac{d}{dr}
-
\frac{l(l+1)}{r^2}
\right)
+
V_{\mathrm{N}}^{(\tau)}(r)
+
\delta_{\tau\mathrm{p}}V_{\mathrm{C}}(r)
&
+
\xi_{lj}V_{\mathrm{SO}}^{(\tau)}(r)
\Bigg]
R_{n_r l j}^{(\tau)}(r)
=
E_{n_r l j}^{(\tau)}
R_{n_r l j}^{(\tau)}(r).
\end{aligned}
\label{eq:radial_tise_background}
\end{equation}

The polar equation is

\begin{equation}
\frac{1}{\sin\theta}
\frac{d}{d\theta}
\left(
\sin\theta
\frac{d\Theta_{l m_l}}{d\theta}
\right)
+
\left[
l(l+1)
-
\frac{m_l^2}{\sin^2\theta}
\right]
\Theta_{l m_l}
=
0,
\label{eq:polar_tise_background}
\end{equation}

and the azimuthal equation is

\begin{equation}
\frac{d^2\Phi_{m_l}}{d\phi^2}
+
m_l^2\Phi_{m_l}
=
0.
\label{eq:azimuthal_tise_background}
\end{equation}

WaveNet uses separate subnetworks to approximate the radial, polar, and
azimuthal components. The formulation is written for general \(m_l\),
while the experiments reported in this work use the representative
choice \(m_l=0\).
The radial component is resolved in fixed \((l,j)\) channels through the
projected spin--orbit eigenvalue \(\xi_{lj}\), whereas the learned polar
and azimuthal components represent scalar orbital functions with
magnetic projection \(m_l\).
\subsection{Physical Constraints}
\label{subsec:physical_constraints_background}

Only the physical constraints that enter the training objective are
summarized here. Their numerical implementation and corresponding loss
terms are presented in Section~\ref{sec:methods}.

For a bound state on the truncated radial domain
\(r\in[0,r_{\max}]\), the radial component satisfies

\begin{equation}
R_{n_r l j}^{(\tau)}(r_{\max})=0.
\label{eq:outer_radial_boundary}
\end{equation}

Regularity at the radial domain origin is imposed through channel-dependent value or
derivative conditions. The precise conditions used for each value of
\(l\) are specified in the methodology. The physical justification for
these conditions is discussed in
Refs.~\cite{khelashvili2011boundary,nistDLMF}.

The azimuthal component is periodic,

\begin{equation}
\Phi_{m_l}(0)=\Phi_{m_l}(2\pi),
\qquad
\Phi_{m_l}'(0)=\Phi_{m_l}'(2\pi).
\label{eq:azimuthal_periodicity}
\end{equation}

Writing the complex azimuthal component as

\begin{equation}
\Phi_{m_l}(\phi)
=
a_{m_l}(\phi)+ib_{m_l}(\phi),
\label{eq:complex_azimuthal_component}
\end{equation}

the phase convention used by the PINN is

\begin{equation}
a_{m_l}(0)=1,
\qquad
b_{m_l}(0)=0,
\qquad
a_{m_l}'(0)=0,
\qquad
b_{m_l}'(0)=m_l.
\label{eq:azimuthal_phase_conditions}
\end{equation}

Before evaluating the physics-informed terms, every separable
wavefunction is normalized explicitly. For state \(k\) of system \(s\),
the component integrals are

\begin{equation*}
I_{R,sk}=\int_{0}^{r_{\max}}\!\left|R_{sk}(r)\right|^2r^2\,dr,
\qquad
I_{\Theta,sk}=\int_{0}^{\pi}\!\left|\Theta_{sk}(\theta)\right|^2
\sin\theta\,d\theta,
\qquad
I_{\Phi,sk}=\int_{0}^{2\pi}\!\left[a_{sk}^{2}(\phi)+b_{sk}^{2}(\phi)\right]d\phi.
\end{equation*}

The normalization scale is

\begin{equation}
\alpha_{sk}
=
\left(
I_{R,sk}
I_{\Theta,sk}
I_{\Phi,sk}
\right)^{-1/2}.
\label{eq:normalization_factor}
\end{equation}

To avoid an arbitrary redistribution of amplitudes among the three
coordinate networks, the scaling factor is applied only to the radial
component,

\begin{equation*}
\widetilde{R}_{sk}(r)
=
\alpha_{sk}R_{sk}(r).
\end{equation*}

The normalized spatial wavefunction is therefore

\begin{equation}
\widetilde{\Psi}_{sk}(r,\theta,\phi)
=
\widetilde{R}_{sk}(r)
\Theta_{sk}(\theta)
\Phi_{sk}(\phi).
\label{eq:normalized_wavefunction}
\end{equation}

The integrals are evaluated by trapezoidal quadrature on fixed radial,
polar, and azimuthal grids. The normalization operation remains inside
the automatic-differentiation graph and is recomputed during every
optimization iteration. Consequently, no additional soft normalization
loss is used.

\subsection{Inferred Woods--Saxon Parameters}
\label{subsec:inferred_parameters_background}

The inverse problem considers one global six-parameter interaction,

\begin{equation}
\bm{\vartheta}
=
\left(
V_0,
\kappa,
r_0,
a,
\lambda_{\mathrm{SO}},
r_{0,\mathrm{SO}}
\right),
\qquad
a_{\mathrm{SO}}=a.
\label{eq:six_parameter_vector}
\end{equation}

These quantities define a shared parameterization rather than
nucleus-specific values. The model therefore learns them jointly from
multiple nucleus--species systems containing different masses, isospin
asymmetries, and spin--orbit splittings.

Given the global spectral dataset

\begin{equation*}
\mathcal{D}
=
\left\{
A_s,
Z_s,
\tau_s,
\left(
E_{s,k}^{\mathrm{obs}},
n_{r,s,k},
l_{s,k},
j_{s,k}
\right)_{k=1}^{K}
\right\}_{s=1}^{N_{\mathrm{sys}}},
\end{equation*}

the dataset-conditioned inverse identification is expressed as

\begin{equation}
\mathcal{D}
\longmapsto
q_{\varphi}
\left(
\bm{\vartheta}
\mid
\mathcal{D}
\right).
\label{eq:dataset_conditioned_identification}
\end{equation}

Here, the network parameters are optimized for the fixed identification
dataset \(\mathcal{D}\). The notation therefore represents
dataset-conditioned inverse identification rather than an amortized
operator for previously unseen spectra.


\section{Methodology}
\label{sec:methods}
The methodology combines three components: an independent
finite-difference forward solver used to generate reference spectra and
perform closure tests, a probabilistic physics-informed neural solver
used for inverse parameter estimation, and an evaluation protocol based
on synthetic and experimental single-particle spectra. The forward
eigenvalue problem is an essential part of the inverse formulation. A
trial Woods--Saxon parameter vector defines a Hamiltonian whose
wavefunctions and Rayleigh energies are evaluated within the
physics-informed objective. The inverse problem is then solved by
adjusting the parameter distribution until this forward response is
consistent with the observed single-particle spectrum and the imposed
physical constraints.
\subsection{Dataset Generation}
\label{subsec:dataset_generation}

Two synthetic datasets of bound single-particle energies were generated
using an independent one-dimensional radial finite-difference solver.
The first dataset was constructed from the Seminole parameterization of
Schwierz et al., while the second was constructed from the Wahlborn
parameterization introduced in
Section~\ref{subsec:woods_saxon_hamiltonian}.

For each nucleus, nucleon species
\(\tau\in\{\mathrm{n},\mathrm{p}\}\), and angular-momentum channel
\((l,j)\), the reduced radial wavefunction was defined as

\begin{equation}
u_{n_rlj}^{(\tau)}(r)
=
rR_{n_rlj}^{(\tau)}(r).
\label{eq:fd_reduced_radial}
\end{equation}

The corresponding radial eigenvalue problem was solved on the finite
domain

\begin{equation*}
0\leq r\leq r_{\max},
\qquad
r_{\max}=25~\mathrm{fm},
\end{equation*}

with homogeneous Dirichlet boundary conditions,

\begin{equation*}
u_{n_rlj}^{(\tau)}(0)
=
u_{n_rlj}^{(\tau)}(r_{\max})
=
0.
\end{equation*}

The radial coordinate was discretized uniformly using \(N=2000\) grid
points and spacing

\begin{equation*}
\Delta r
=
\frac{r_{\max}}{N-1}.
\end{equation*}

At each interior point, the second derivative was approximated by

\begin{equation}
\left.
\frac{d^2u}{dr^2}
\right|_{r=r_i}
\approx
\frac{u_{i+1}-2u_i+u_{i-1}}{(\Delta r)^2},
\qquad
\Delta r=\frac{r_{\max}}{N-1}.
\label{eq:fd_second_derivative}
\end{equation}

The resulting discretized eigenvalue problem is

\begin{equation}
\bm{H}_{\tau lj}\bm{u}_k
=
E_k\bm{u}_k,
\qquad
K_\tau
=
\frac{(\hbar c)^2}{2\mu_\tau},
\label{eq:dataset_matrix_eigenproblem}
\end{equation}

where \(\bm{H}_{\tau lj}\) is a real symmetric tridiagonal matrix. Its nonzero elements are

\begin{equation}
\left(\bm{H}_{\tau lj}\right)_{ii}
=
\frac{2K_\tau}{(\Delta r)^2}
+
V_{\tau lj}^{\mathrm{eff}}(r_i),
\qquad
\left(\bm{H}_{\tau lj}\right)_{i,i\pm1}
=
-\frac{K_\tau}{(\Delta r)^2}.
\label{eq:fd_hamiltonian_elements}
\end{equation}

The lowest eigenvalues were calculated using the SciPy sparse eigensolver
\texttt{eigsh}, configured to return the smallest algebraic eigenvalues.
Only bound states satisfying \(E_k<0\) are retained. Within each \((l,j)\) channel, the bound energies were
ordered from the most deeply bound state upward and assigned
\(n_r=0,1,\ldots\). Channels up to \(l_{\max}=5\) and radial quantum
numbers up to \(n_{r,\max}=3\) were considered.

The parameters used to generate the synthetic datasets are reported in
Table~\ref{tab:dataset_parameterizations}.

\begin{table}[H]
\centering
\caption{
Woods--Saxon parameters used to generate the synthetic
single-particle spectra.
}
\label{tab:dataset_parameterizations}
\small
\setlength{\tabcolsep}{4.5pt}
\renewcommand{\arraystretch}{1.08}
\begin{tabular}{lccccccc}
\toprule
Parameterization
& \(V_0\) [MeV]
& \(\kappa\)
& \(r_0\) [fm]
& \(a\) [fm]
& \(\lambda_{\mathrm{SO}}\)
& \(r_{0,\mathrm{SO}}\) [fm]
& \(a_{\mathrm{SO}}\) [fm]
\\
\midrule
Seminole \(\bigl(\bm{\vartheta}_{\mathrm{S}}^{\mathrm{ref}}\bigr)\)
& 52.06
& 0.639
& 1.260
& 0.662
& 24.1
& 1.160
& 0.662
\\
Wahlborn \(\bigl(\bm{\vartheta}_{\mathrm{W}}^{\mathrm{ref}}\bigr)\)
& 51.00
& 0.670
& 1.270
& 0.670
& 32.0
& 1.270
& 0.670
\\
\bottomrule
\end{tabular}
\end{table}

The complete datasets contain neutron and proton spectra for
\(^{12}\mathrm{C}\), \(^{16}\mathrm{O}\), \(^{20}\mathrm{Ne}\),
\(^{24}\mathrm{Mg}\), \(^{28}\mathrm{Si}\), \(^{32}\mathrm{S}\),
\(^{40}\mathrm{Ca}\), \(^{48}\mathrm{Ca}\), \(^{56}\mathrm{Ni}\),
\(^{40}\mathrm{Ar}\), \(^{48}\mathrm{Ti}\), \(^{60}\mathrm{Ni}\),
\(^{72}\mathrm{Zn}\), \(^{90}\mathrm{Zr}\),
\(^{100}\mathrm{Sn}\), \(^{132}\mathrm{Sn}\), and
\(^{208}\mathrm{Pb}\). Each dataset therefore contains \(34\)
nucleus--species systems.

A system is represented as

\begin{equation}
\mathcal{D}_{s}
=
\left\{
A_s,\,
Z_s,\,
\tau_s,\,
\mathcal{S}_{s}
\right\},
\label{eq:dataset_system}
\end{equation}

where

\begin{equation}
\mathcal{S}_{s}
=
\left\{
\left(
n_{r,sk},
l_{sk},
j_{sk},
E_{sk}^{\mathrm{obs}}
\right)
\right\}_{k=1}^{K_s}
\label{eq:dataset_states}
\end{equation}

contains the bound states associated with that system.

For each inverse experiment, the same number \(K\) of states was retained
from every selected system. The pair-preserving selection procedure first
included the most deeply bound state, then prioritized complete
spin--orbit doublets sharing \((n_r,l)\), and finally included the
remaining bound states until \(K\) states were obtained. Systems without
the required number of physical states were excluded; no artificial
states were introduced.

\subsection{Synthetic Global Parameter Identification}
\label{subsec:synthetic_parameter_identification}

The global inverse solver was first evaluated on synthetic spectra
generated from the Seminole and Wahlborn parameterizations defined in
Section~\ref{subsec:woods_saxon_hamiltonian}. Because the generating
parameters are known, these experiments provide a controlled test of
six-parameter recovery, with \(a_{\mathrm{SO}}=a\) and the reference parameter vectors were \(\bm{\vartheta}_{\mathrm{S}}^{\mathrm{ref}}\) and \(\bm{\vartheta}_{\mathrm{W}}^{\mathrm{ref}}\) from Table~\ref{tab:dataset_parameterizations}.

\subsubsection{Synthetic Datasets}
\label{subsubsec:synthetic_identification_datasets}

Both experiments used spectra from \(^{40}\mathrm{Ca}\),
\(^{48}\mathrm{Ca}\), \(^{132}\mathrm{Sn}\), and
\(^{208}\mathrm{Pb}\), with seven states selected from each
nucleus--species system. The selection spans deeply and weakly bound
levels and includes spin--orbit doublets, thereby sampling the bulk and
surface regions of the potential. Variation in nuclear mass and isospin
asymmetry additionally helps separate \(V_0\) from \(\kappa\).

For Wahlborn, both proton and neutron spectra were included to test the
shared-geometry formulation across both species. Other established
Woods--Saxon parameterizations have instead used species-dependent
values of \(r_0\), \(a\), and \(r_{0,\mathrm{SO}}\)
\cite{rost1968proton,dudek1979parameters,dudek1982description}.

The Seminole parameterization already defines a unified proton--neutron
interaction with shared geometry and an explicit isospin-dependent
central depth. Neutron spectra spanning several asymmetries were
therefore used for its recovery, while both species were retained for
the subsequent closure test.

The selected synthetic states and their finite-difference energies are
reported in Appendix~\ref{app:synthetic_states}.

\subsubsection{Experimental Dataset}
\label{subsubsec:experimental_configuration}

The datasets used for the experimental procedure were a reduced dataset for parameter
identification and a broader dataset for spectral evaluation.

The parameter-identification dataset was constructed from measured
neutron and proton single-particle energies in

\begin{equation*}
^{40}\mathrm{Ca},
\qquad
^{48}\mathrm{Ca},
\qquad
^{132}\mathrm{Sn},
\qquad
^{208}\mathrm{Pb}.
\end{equation*}

Six states were selected from each nucleus--species system so that all
system-level context vectors had the same dimensionality. The
\(^{40}\mathrm{Ca}\) proton system was excluded because only four
experimental bound states were available.

The final identification dataset contains seven nucleus--species
systems and \(42\) experimental levels. The neutron subset comprises
\(24\) levels from all four nuclei, while the proton subset comprises
\(18\) levels from \(^{48}\mathrm{Ca}\),
\(^{132}\mathrm{Sn}\), and \(^{208}\mathrm{Pb}\). The selected states
cover both deeply and weakly bound orbitals and include \(14\) complete
spin--orbit doublets, with experimental splitting magnitudes between
approximately \(0.52\) and \(6.80~\mathrm{MeV}\).

The same nuclei, nucleon species, quantum-number assignments, and
experimental energies were used for the Seminole and Wahlborn
experiments. Differences between the inferred interactions therefore
originate from the potential expressions rather than from different
identification datasets.

A broader dataset was used for the final spectral evaluation. It
contains \(96\) experimental entries associated with

\begin{equation*}
^{16}\mathrm{O},
\quad
^{40}\mathrm{Ca},
\quad
^{48}\mathrm{Ca},
\quad
^{56}\mathrm{Ni},
\quad
^{90}\mathrm{Zr},
\quad
^{100}\mathrm{Sn},
\quad
^{132}\mathrm{Sn},
\quad
^{208}\mathrm{Pb}.
\end{equation*}

For all nuclei except \(^{90}\mathrm{Zr}\), the energies and
quantum-number assignments were taken from the experimental compilation
used by Schwierz et al.\
\cite{schwierz2007parameterization}. The reference Seminole parameters
were obtained in that work through a global least-squares fit to
single-particle and single-hole spectra around

\begin{equation*}
^{16}\mathrm{O},
\quad
^{40}\mathrm{Ca},
\quad
^{48}\mathrm{Ca},
\quad
^{56}\mathrm{Ni},
\quad
^{100}\mathrm{Sn},
\quad
^{132}\mathrm{Sn},
\quad
^{208}\mathrm{Pb}.
\end{equation*}

The \(^{90}\mathrm{Zr}\) entries were added as an independent
evaluation case. The neutron-hole information below the \(N=50\) shell
closure was obtained from
\(^{90}\mathrm{Zr}(p,d)^{89}\mathrm{Zr}\) measurements
\cite{ball1968neutron,kasagi1983highly,duhamel1991neutron}.
The neutron-particle information above the shell closure was obtained
from \(^{90}\mathrm{Zr}(d,p)^{91}\mathrm{Zr}\) and
\(^{90}\mathrm{Zr}(\alpha,{}^{3}\mathrm{He})^{91}\mathrm{Zr}\)
measurements
\cite{bingham1970neutron,graue1972highresolution,sharp2012trends}.
Where the spectroscopic strength of an orbital was distributed over
several measured states, the adopted energy corresponds to the
spectroscopic-strength-weighted centroid.

The evaluation nuclei consequently represent three different cases.
The nuclei \(^{40}\mathrm{Ca}\), \(^{48}\mathrm{Ca}\),
\(^{132}\mathrm{Sn}\), and \(^{208}\mathrm{Pb}\) contributed to both
the PINN identification dataset and the least-squares calibration of the
reference Seminole interaction. The nuclei
\(^{16}\mathrm{O}\), \(^{56}\mathrm{Ni}\), and
\(^{100}\mathrm{Sn}\) were excluded from PINN training but contributed
to the reference calibration. The \(^{90}\mathrm{Zr}\) spectrum
contributed to neither parameter-determination procedure and therefore
provides an out-of-calibration test for both interactions.

Experimental and calculated levels were matched according to

\begin{equation}
\left(
A,Z,\tau,n_r,l,j
\right),
\label{eq:experimental_state_matching}
\end{equation}

where \(\tau\) denotes the nucleon species. For the direct comparison of
the reference and PINN-inferred Seminole interactions, a common
state-selection mask was applied. Only experimental levels with bound
calculated energies in both spectra were included.

Two of the \(96\) experimental entries were excluded by this condition.
The \(^{132}\mathrm{Sn}\) neutron \(1h_{9/2}\) orbital is not bound for
the PINN-inferred parameter set, while the \(^{208}\mathrm{Pb}\) proton
\(3p_{1/2}\) orbital is not bound in either Seminole spectrum. The
resulting common Seminole benchmark contains \(94\) levels, comprising
\(59\) neutron and \(35\) proton states.

Common state-selection masks were applied separately to the Seminole
and Wahlborn comparisons. For Seminole, the common \(94\)-state
benchmark is used for both the aggregate results in
Table~\ref{tab:experimental_spectral_metrics} and the nucleus-resolved
results in Table~\ref{tab:seminole_performance_by_nucleus}. For
Wahlborn, one common mask is applied to the reference,
distribution-mean, and best sampled interactions. Thus, all
within-expression error comparisons are evaluated on identical sets of
experimental levels.

In Table~\ref{tab:seminole_performance_by_nucleus}, a dagger denotes a
nucleus excluded from PINN training. A double dagger additionally
denotes a nucleus excluded from the least-squares calibration of the
reference Seminole interaction.

\subsection{Finite-Difference Least-Squares Fit}
\label{subsec:fd_lsq_fit}

To provide a conventional baseline under the same data conditions as the
physics-informed model, the six Woods--Saxon parameters defined in
Eq.~\eqref{eq:six_parameter_vector} were also estimated by nonlinear
least squares. Following the fitting logic of the Seminole calibration,
each trial parameter vector was evaluated with the finite-difference
radial solver of Sec.~\ref{subsec:dataset_generation}, and the resulting energies
were matched to the same \(42\) experimental levels used for PINN
identification according to Eq.~\eqref{eq:experimental_state_matching}.

The fitted parameters minimized
\begin{equation}
\chi^{2}(\bm{\vartheta})
=
\sum_{k=1}^{42}
\left[
E_{k}^{\mathrm{FD}}(\bm{\vartheta})
-
E_{k}^{\mathrm{obs}}
\right]^{2},
\label{eq:fd_lsq_objective}
\end{equation}
using the unconstrained Levenberg--Marquardt algorithm
\cite{levenberg1944method,marquardt1963algorithm}. Initial parameter
vectors were sampled uniformly from physically motivated intervals.
The converged solution with the smallest
\(\chi^{2}\) was retained and evaluated without further adjustment on
the broader experimental benchmark.

\subsection{Physics-Informed Neural Solver}
\label{subsec:physics_informed_solver}

\subsubsection{Architecture Overview}
\label{subsubsec:architecture_overview}

The proposed solver couples a conditional wavefunction network, denoted
WaveNet, with a global probabilistic parameter-inference network, denoted
ParamNet. The input to an inverse experiment is the collection

\begin{equation}
\mathcal{D}
=
\left\{
\mathcal{D}_{s}
\right\}_{s=1}^{N_{\mathrm{sys}}},
\label{eq:global_dataset}
\end{equation}

where each \(\mathcal{D}_{s}\) represents one nucleus--species system
and contains its nuclear descriptors, selected quantum states, and
single-particle energies.

WaveNet is shared across all systems and reconstructs the separated
radial, polar, and azimuthal components of every selected state. After
explicit differentiable normalization, the sampled radial functions are
combined with the corresponding energies, nuclear descriptors, and
quantum numbers to construct one fixed-dimensional context vector for
each nucleus--species system.

ParamNet applies a shared encoder to all system-level context vectors.
The encoded representations are aggregated through permutation-invariant
mean pooling and mapped to one global variational distribution over the
six Woods--Saxon parameters defined in Eq.~\eqref{eq:six_parameter_vector}.

The complete computational sequence is

\begin{equation}
\begin{aligned}
\mathcal{D}
&\longrightarrow
\mathrm{WaveNet}
\longrightarrow
\left\{
\widetilde{R}_{sk},
\Theta_{sk},
\Phi_{sk}
\right\}_{s,k}
&\longrightarrow
\left\{
\bm{x}_{s}
\right\}_{s=1}^{N_{\mathrm{sys}}}
\longrightarrow
\mathrm{mean\ pooling}
\longrightarrow
q_{\varphi}
\left(
\bm{z}\mid\mathcal{D}
\right)
\\
&\longrightarrow
\bm{\vartheta}
\longrightarrow
\left\{
\widehat{H}_{\tau_s}
\left(
\bm{\vartheta}
\right)
\right\}_{s=1}^{N_{\mathrm{sys}}}.
\end{aligned}
\label{eq:solver_workflow}
\end{equation}

A single latent sample is transformed into one physical parameter vector
and used consistently in the Hamiltonians of all systems contributing to
the optimization step. The framework therefore identifies one global interaction for the
selected dataset rather than separate Woods--Saxon parameters for each
nucleus. WaveNet and ParamNet are retrained when the identification
dataset changes; no amortized inference on a new dataset is assumed.

\begin{center}
\centering
\includegraphics[width=0.88\linewidth]{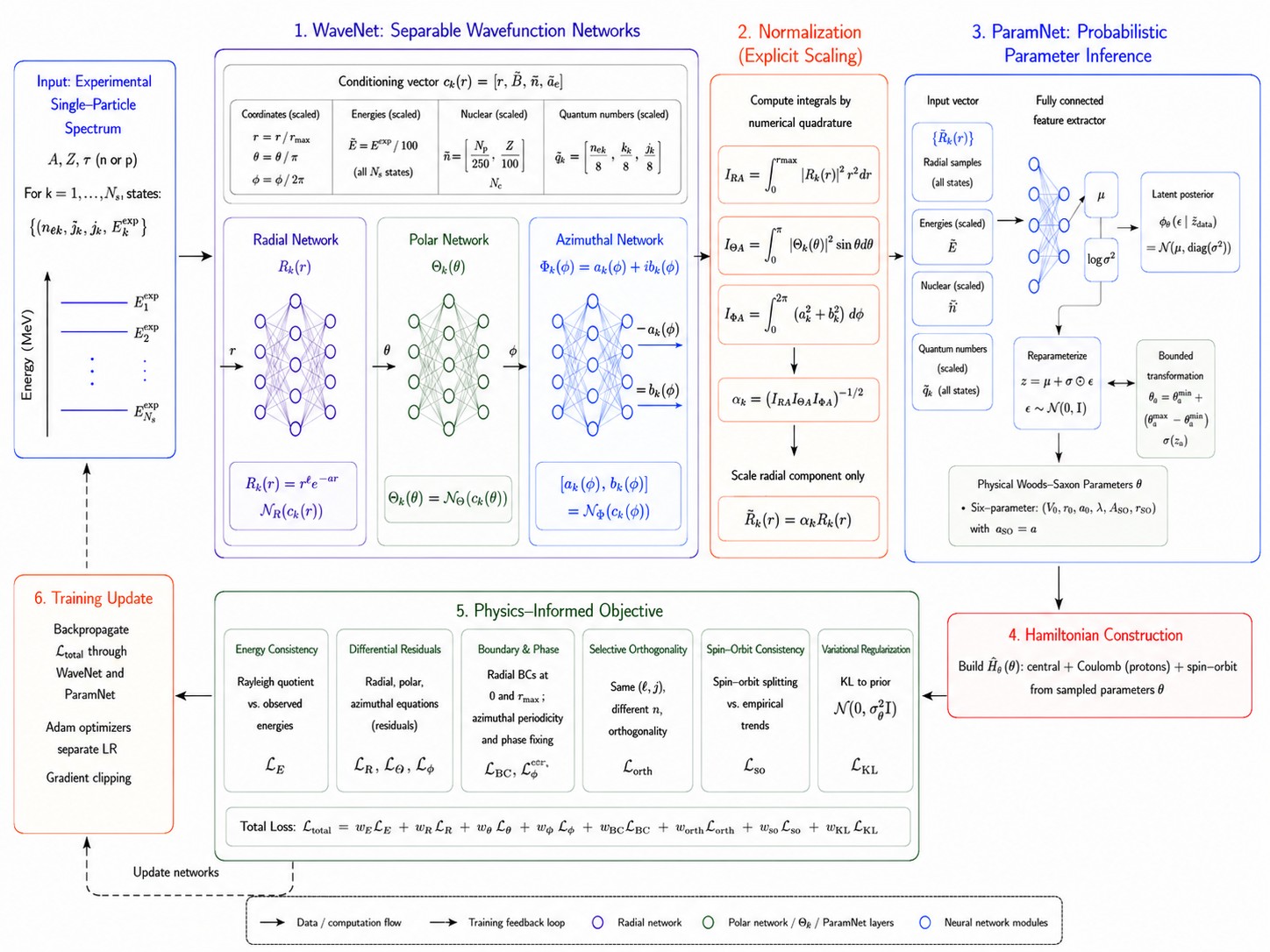}
\captionof{figure}{
Architecture of the global probabilistic physics-informed solver.
WaveNet reconstructs the separated wavefunction components, which are
normalized and used to form one context vector per nucleus--species
system. ParamNet mean-pools the encoded system contexts to construct one
global six-parameter variational distribution for the current
identification dataset. Samples from this distribution define the shared
Woods--Saxon Hamiltonian. The resulting spectral and physics-informed
losses jointly update WaveNet and ParamNet.
}
\label{fig:pinn_architecture}
\end{center}

\subsubsection{Conditional WaveNet}
\label{subsubsec:wavenet}

WaveNet contains three coordinate-specific multilayer perceptrons,
\(\mathcal{N}_{R}\), \(\mathcal{N}_{\Theta}\), and
\(\mathcal{N}_{\Phi}\), which approximate the radial, polar, and
azimuthal components, respectively. The same three networks are shared
across all selected states and nuclear systems.

For state \(k\) of system \(s\), the conditioning vector associated with
coordinate \(x\in\{r,\theta,\phi\}\) is

\begin{equation}
\bm{c}_{sk}(x)
=
\left[
\overline{x},\,
\overline{\bm{E}}_{s},\,
\overline{\bm{n}}_{s},\,
\overline{\bm{q}}_{sk}
\right].
\label{eq:wavenet_context}
\end{equation}

The spectrum, nuclear descriptors, state descriptors, and coordinates
are scaled as

\begin{equation*}
\begin{aligned}
\overline{\bm{E}}_{s}
&=
\frac{1}{100}
\left[E_{s1}^{\mathrm{obs}},\ldots,E_{sK}^{\mathrm{obs}}\right],
&
\overline{\bm{n}}_{s}
&=
\left[\frac{A_s}{250},\frac{Z_s}{100},
\mathbb{I}_{\tau_s=\mathrm{p}}\right],
\\
\overline{\bm{q}}_{sk}
&=
\left[\frac{n_{r,sk}}{5},\frac{l_{sk}}{8},\frac{j_{sk}}{8}\right],
&
(\overline{r},\overline{\theta},\overline{\phi})
&=
\left(\frac{r}{r_{\max}},\frac{\theta}{\pi},\frac{\phi}{2\pi}\right).
\end{aligned}
\end{equation*}

The radial output is represented as

\begin{equation}
R_{sk}(r)
=
r^{l_{sk}}
\exp(-\beta r)
\mathcal{N}_{R}
\left(
\bm{c}_{sk}(r)
\right).
\label{eq:wavenet_radial_ansatz}
\end{equation}

The factor \(r^{l_{sk}}\) incorporates the expected regular behavior near
the origin, while the exponential factor promotes decay at large radius.
The neural network learns the remaining state-dependent radial
structure.

The polar component is

\begin{equation}
\Theta_{sk}(\theta)
=
\mathcal{N}_{\Theta}
\left(
\bm{c}_{sk}(\theta)
\right),
\label{eq:wavenet_theta}
\end{equation}

whereas the azimuthal network returns two outputs corresponding to the
real and imaginary components,

\begin{equation}
\Phi_{sk}(\phi)
=
a_{sk}(\phi)
+
ib_{sk}(\phi),
\label{eq:wavenet_phi}
\end{equation}

with

\begin{equation*}
\left[
a_{sk}(\phi),
b_{sk}(\phi)
\right]
=
\mathcal{N}_{\Phi}
\left(
\bm{c}_{sk}(\phi)
\right).
\end{equation*}

All subnetworks use \(\tanh\) activations. Linear weights are
initialized using Xavier initialization and biases are initialized to
zero. Spatial derivatives required by the governing equations are
computed through automatic differentiation.

\subsubsection{Probabilistic ParamNet}

ParamNet receives one fixed-dimensional context vector from each nucleus--species system. For system $s$, the normalized radial wavefunction of each selected state is evaluated at $N_p$ uniformly spaced radial probe points,
\[
\boldsymbol{r}_p = [r_1,\ldots,r_{N_p}] .
\]

The sampled radial representation is
\[
\widetilde{\boldsymbol{R}}_{sk}
=
\left[
\widetilde{R}_{sk}(r_1),\ldots,\widetilde{R}_{sk}(r_{N_p})
\right] .
\]

Because eigenfunctions are defined only up to an overall sign, a deterministic sign convention is applied to every sampled radial vector using a fixed interior probe point. This removes an arbitrary input ambiguity without changing any physical observable.

The complete system-level context is
\begin{equation}
\boldsymbol{x}_s
=
\left[
\widetilde{\boldsymbol{R}}_{s1},\ldots,
\widetilde{\boldsymbol{R}}_{sK},
\boldsymbol{E}_s,
\bar{n}_s,
\bar{\boldsymbol{q}}_{s1},\ldots,
\bar{\boldsymbol{q}}_{sK}
\right] .
\tag{35}
\end{equation}

with dimensionality

\begin{equation*}
d_{\mathrm{in}}
=
KN_p
+
K
+
3
+
3K.
\end{equation*}

Each system context is processed by the same encoder,

\begin{equation}
\bm{h}_{s}
=
f_{\mathrm{enc}}
\left(
\bm{x}_{s}
\right),
\label{eq:paramnet_system_encoding}
\end{equation}

and the encoded system features are combined through
permutation-invariant mean pooling,

\begin{equation}
\bm{h}_{\mathrm{global}}
=
\frac{1}{N_{\mathrm{sys}}}
\sum_{s=1}^{N_{\mathrm{sys}}}
\bm{h}_{s}.
\label{eq:paramnet_global_pooling}
\end{equation}

Two output heads predict the latent mean and log-variance,

\begin{equation*}
\bm{\mu}
=
f_{\mu}
\left(
\bm{h}_{\mathrm{global}}
\right),
\qquad
\log\bm{\sigma}^{2}
=
f_{\sigma}
\left(
\bm{h}_{\mathrm{global}}
\right).
\end{equation*}

The resulting six-dimensional latent distribution is

\begin{equation}
q_{\varphi}
\left(
\bm{z}\mid\mathcal{D}
\right)
=
\mathcal{N}
\left(
\bm{z};
\bm{\mu},
\operatorname{diag}
\left(
\bm{\sigma}^{2}
\right)
\right),
\label{eq:paramnet_variational_distribution}
\end{equation}

where

\begin{equation*}
\bm{\sigma}
=
\exp
\left(
\frac{1}{2}
\log\bm{\sigma}^{2}
\right).
\end{equation*}

During training, a latent sample is generated using the
reparameterization trick \cite{kingma2014autoencoding},

\begin{equation}
\bm{z}
=
\bm{\mu}
+
\bm{\sigma}\odot\bm{\epsilon},
\qquad
\bm{\epsilon}
\sim
\mathcal{N}
\left(
\bm{0},
\bm{I}
\right).
\label{eq:reparameterization_trick}
\end{equation}

By isolating the stochasticity in \(\bm{\epsilon}\), this formulation
allows gradients to propagate through the sampled latent variable to
\(\bm{\mu}\) and \(\bm{\sigma}\), generally reducing gradient-estimation
variance and improving optimization stability.

Each latent component is mapped to a bounded physical parameter through

\begin{equation}
\vartheta_i
=
\vartheta_i^{\min}
+
\left(
\vartheta_i^{\max}
-
\vartheta_i^{\min}
\right)
\operatorname{sigmoid}(z_i).
\label{eq:bounded_parameter_mapping}
\end{equation}

The admissible parameter intervals are

\begin{equation*}
\begin{aligned}
40
&\leq
V_0
\leq
65~\mathrm{MeV},
&
0.30
&\leq
\kappa
\leq
1.00,
\\
1.15
&\leq
r_0
\leq
1.35~\mathrm{fm},
&
0.55
&\leq
a
\leq
0.75~\mathrm{fm},
\\
15
&\leq
\lambda_{\mathrm{SO}}
\leq
40,
&
0.90
&\leq
r_{0,\mathrm{SO}}
\leq
1.35~\mathrm{fm}.
\end{aligned}
\end{equation*}

These intervals cover the ranges of central, isospin-dependent,
geometric, and spin--orbit parameters reported for the Rost, Chepurnov,
Wahlborn, Universal, Optimized, and Seminole Woods--Saxon
parameterizations
\cite{rost1968proton,chepurnov1968average,dudek1979parameters,
dudek1982description,schwierz2007parameterization}. The search domain is
therefore physically informed while remaining sufficiently broad to
avoid restricting ParamNet to the neighborhood of either synthetic
reference interaction.

The spin--orbit diffuseness is not inferred independently but is
constrained according to
\(a_{\mathrm{SO}}=a\). At inference, multiple samples are drawn from
the learned latent distribution and transformed into physical parameter
space without additional optimization. Their physical-space mean
defines the primary, selection-free parameter estimate, while their
standard deviations define the learned output spread. Individual
samples represent realizations of the parameter region inferred from
the identification dataset. For the experimental comparison, the
sampled set yielding the lowest spectral MAE on the broader
benchmark is additionally reported as the best sampled PINN
interaction, separately from the distribution-mean estimator. Within
the fixed model and training procedure, smaller spreads are interpreted
as greater model-derived confidence and stronger qualitative parameter
stiffness; larger spreads indicate greater admissible flexibility or
flatter effective loss directions. More informative spectra are
therefore expected to reduce the spread of the parameters they constrain
most strongly. These standard deviations are not calibrated Bayesian
credible intervals.
\subsection{Physics-Informed Objective}
\label{subsec:physics_informed_objective}

The total objective combines spectral consistency, the separated
Schrödinger equations, boundary and phase conditions, selective
orthogonality, spin--orbit splitting consistency, and latent
regularization.

\subsubsection{Energy Consistency}
\label{subsubsec:energy_consistency}

For a sampled global parameter vector \(\bm{\vartheta}\), the predicted
energy is calculated using the Rayleigh quotient,

\begin{equation}
E_{sk}^{\mathrm{RQ}}
=
\frac{
\left\langle
\widetilde{\Psi}_{sk}
\middle|
\widehat{H}_{\tau_s}
\left(
\bm{\vartheta}
\right)
\middle|
\widetilde{\Psi}_{sk}
\right\rangle
}{
\left\langle
\widetilde{\Psi}_{sk}
\middle|
\widetilde{\Psi}_{sk}
\right\rangle
}.
\label{eq:rayleigh_energy}
\end{equation}

The three-dimensional integrals are factorized into radial, polar, and
azimuthal contributions and evaluated using numerical quadrature. For a
batch \(\mathcal{B}\) of systems, the energy loss is

\begin{equation}
\mathcal{L}_{E}
=
\frac{1}{
\sum_{s\in\mathcal{B}}K_s
}
\sum_{s\in\mathcal{B}}
\sum_{k=1}^{K_s}
\left(
E_{sk}^{\mathrm{RQ}}
-
E_{sk}^{\mathrm{obs}}
\right)^2.
\label{eq:energy_loss}
\end{equation}

\subsubsection{Differential-Equation Residuals}
\label{subsubsec:differential_residuals}

The radial, polar, and azimuthal residuals are evaluated at randomly
sampled collocation points. Derivatives are calculated through automatic
differentiation.

To remove explicit \(1/r\) and \(1/r^2\) singularities, the radial
equation is multiplied by \(r^2\). Defining

\begin{equation*}
V_{sk}^{\mathrm{eff}}
\left(
r;\bm{\vartheta}
\right)
=
V_{\mathrm{N}}^{(\tau_s)}(r)
+
\delta_{\tau_s\mathrm{p}}V_{\mathrm{C}}(r)
+
\xi_{l_{sk}j_{sk}}
V_{\mathrm{SO}}^{(\tau_s)}(r),
\end{equation*}

the radial residual is

\begin{equation}
\begin{aligned}
\mathcal{R}_{R,sk}(r)
={}&
-K_{\tau_s}
\left[
r^2
\widetilde{R}_{sk}''(r)
+
2r
\widetilde{R}_{sk}'(r)
-
l_{sk}(l_{sk}+1)
\widetilde{R}_{sk}(r)
\right]
+
r^2
\left[
V_{sk}^{\mathrm{eff}}
\left(
r;\bm{\vartheta}
\right)
-
E_{sk}^{\mathrm{RQ}}
\right]
\widetilde{R}_{sk}(r).
\end{aligned}
\label{eq:radial_residual}
\end{equation}

The corresponding loss is

\begin{equation}
\mathcal{L}_{R}
=
\frac{1}{
\sum_{s\in\mathcal{B}}K_sN_R
}
\sum_{s\in\mathcal{B}}
\sum_{k=1}^{K_s}
\sum_{i=1}^{N_R}
\left|
\mathcal{R}_{R,sk}(r_i)
\right|^2.
\label{eq:radial_residual_loss}
\end{equation}

The polar equation is evaluated in the scaled form

\begin{equation}
\begin{aligned}
\mathcal{R}_{\Theta,sk}(\theta)
={}&
\sin^2\theta\,
\Theta_{sk}''(\theta)
+
\sin\theta\cos\theta\,
\Theta_{sk}'(\theta)
+
\left[
l_{sk}(l_{sk}+1)\sin^2\theta
-
m_l^2
\right]
\Theta_{sk}(\theta).
\end{aligned}
\label{eq:theta_residual}
\end{equation}

The polar residual loss is

\begin{equation}
\mathcal{L}_{\Theta}
=
\frac{1}{
\sum_{s\in\mathcal{B}}K_sN_{\Theta}
}
\sum_{s\in\mathcal{B}}
\sum_{k=1}^{K_s}
\sum_{i=1}^{N_{\Theta}}
\left|
\mathcal{R}_{\Theta,sk}(\theta_i)
\right|^2.
\label{eq:theta_residual_loss}
\end{equation}

Writing

\begin{equation*}
\Phi_{sk}(\phi)
=
a_{sk}(\phi)
+
ib_{sk}(\phi),
\end{equation*}

the two azimuthal differential residuals are

\begin{equation*}
\mathcal{R}_{\Phi,sk}^{(a)}
=
a_{sk}''+m_l^2a_{sk},
\qquad
\mathcal{R}_{\Phi,sk}^{(b)}
=
b_{sk}''+m_l^2b_{sk}.
\end{equation*}

Their mean-squared contribution is denoted by
\(\mathcal{L}_{\Phi,\mathrm{ODE}}\). The calculations reported in this
work use the representative magnetic projection \(m_l=0\), unless stated
otherwise. Since the Hamiltonian is rotationally invariant, this choice
does not change the single-particle energies.

\subsubsection{Boundary, Periodicity, and Phase Constraints}
\label{subsubsec:boundary_constraints}

The normalized radial solution satisfies the outer boundary condition

\begin{equation*}
\widetilde{R}_{sk}(r_{\max})=0.
\end{equation*}

At the origin, the implemented condition depends on \(l_{sk}\),

\begin{equation}
\mathcal{B}_{0,sk}
=
\begin{cases}
\left|
\widetilde{R}_{sk}'(0)
\right|^2,
&
l_{sk}=0,
\\[2mm]
\left|
\widetilde{R}_{sk}(0)
\right|^2,
&
l_{sk}=1,
\\[2mm]
\left|
\widetilde{R}_{sk}(0)
\right|^2
+
\left|
\widetilde{R}_{sk}'(0)
\right|^2,
&
l_{sk}\geq2.
\end{cases}
\label{eq:radial_origin_conditions}
\end{equation}

The radial boundary loss is

\begin{equation}
\mathcal{L}_{\mathrm{BC}}
=
\frac{1}{
\sum_{s\in\mathcal{B}}K_s
}
\sum_{s\in\mathcal{B}}
\sum_{k=1}^{K_s}
\left[
\left|
\widetilde{R}_{sk}(r_{\max})
\right|^2
+
\mathcal{B}_{0,sk}
\right].
\label{eq:boundary_loss}
\end{equation}

No separate polar boundary penalty is used. Instead, the exact poles are
excluded from the collocation set and the \(\sin^2\theta\)-scaled
residual controls the solution near the coordinate singularities.

The azimuthal function is required to be periodic in both value and
first derivative for both \(a_{sk}\) and \(b_{sk}\). Periodicity alone
does not select the phase convention
\(\Phi_{m_l}(\phi)=e^{im_l\phi}\); therefore, additional phase-fixing
conditions are imposed.

\begin{equation*}
a_{sk}(0)=1,
\qquad
b_{sk}(0)=0,
\qquad
a_{sk}'(0)=0,
\qquad
b_{sk}'(0)=m.
\end{equation*}

The complete azimuthal loss is

\begin{equation}
\mathcal{L}_{\Phi}
=
\mathcal{L}_{\Phi,\mathrm{ODE}}
+
\mathcal{L}_{\Phi,\mathrm{per}}
+
2\mathcal{L}_{\Phi,\mathrm{phase}}.
\label{eq:complete_phi_loss}
\end{equation}

\subsubsection{Orthogonality}
\label{subsubsec:orthogonality}

Explicit orthogonality is imposed only between different radial
excitations belonging to the same \((l,j)\) channel. The penalized pair
set is

\begin{equation}
\mathcal{P}_{\mathrm{orth}}
=
\left\{
(sk,sq):
l_{sk}=l_{sq},\,
j_{sk}=j_{sq},\,
n_{r,sk}\neq n_{r,sq}
\right\}.
\label{eq:orthogonal_pair_set}
\end{equation}

For each eligible pair, the complex overlap is

\begin{equation}
O_{kq}^{(s)}
=
\int_{\Omega}
\widetilde{\Psi}_{sq}^{*}
\widetilde{\Psi}_{sk}
\,d^3r.
\label{eq:wavefunction_overlap}
\end{equation}

The integral is estimated through Monte Carlo sampling in
\((r,\theta,\phi)\), including the spherical Jacobian
\(r^2\sin\theta\). The orthogonality loss is

\begin{equation}
\mathcal{L}_{\mathrm{orth}}
=
\frac{1}{
\left|
\mathcal{P}_{\mathrm{orth}}
\right|
}
\sum_{(sk,sq)\in\mathcal{P}_{\mathrm{orth}}}
\left|
O_{kq}^{(s)}
\right|^2.
\label{eq:orthogonality_loss}
\end{equation}

States with different \(l\) or \(j\), including spin--orbit partners,
are not explicitly penalized because their orthogonality follows from
the spin-angular part of the full physical wavefunction.

\subsubsection{Spin--Orbit Splitting Consistency}
\label{subsubsec:spin_orbit_loss}

To provide an additional direct constraint on the spin--orbit sector,
the predicted and observed splittings of complete spin--orbit doublets
are compared. For each pair sharing \((n_r,l)\) with
\(j_{\pm}=l\pm1/2\), define

\begin{equation*}
\Delta E_{sl}^{\mathrm{RQ}}
=
E_{s,n_rlj_{+}}^{\mathrm{RQ}}
-
E_{s,n_rlj_{-}}^{\mathrm{RQ}},
\end{equation*}

and

\begin{equation*}
\Delta E_{sl}^{\mathrm{obs}}
=
E_{s,n_rlj_{+}}^{\mathrm{obs}}
-
E_{s,n_rlj_{-}}^{\mathrm{obs}}.
\end{equation*}

The splitting loss is

\begin{equation}
\mathcal{L}_{\mathrm{SO}}
=
\frac{1}{
\left|
\mathcal{P}_{\mathrm{SO}}
\right|
}
\sum_{(s,n_r,l)\in\mathcal{P}_{\mathrm{SO}}}
\left(
\Delta E_{sl}^{\mathrm{RQ}}
-
\Delta E_{sl}^{\mathrm{obs}}
\right)^2,
\label{eq:spin_orbit_splitting_loss}
\end{equation}

where \(\mathcal{P}_{\mathrm{SO}}\) contains the complete doublets
available in the selected spectra.

\subsubsection{Variational Regularization}
\label{subsubsec:variational_regularization}

The learned latent distribution is regularized against the isotropic
Gaussian prior $p(\bm{z}) = \mathcal{N} \left(\bm{0},\sigma_p^2\bm{I}\right)$.
For the diagonal variational distribution in
Eq.~\eqref{eq:paramnet_variational_distribution}, the KL divergence is

\begin{equation}
\mathcal{L}_{\mathrm{KL}}
=
\frac{1}{2}
\sum_{i=1}^{6}
\left[
\frac{
\sigma_i^2+\mu_i^2
}{
\sigma_p^2
}
-
1
+
2\log\sigma_p
-
\log\sigma_i^2
\right].
\label{eq:kl_loss}
\end{equation}

This regularization acts in latent space. Because the physical
parameters are obtained through a nonlinear bounded transformation, the
prior should not be interpreted as a Gaussian distribution centered at
the midpoints of the physical intervals. The parameters are inferred
indirectly from the observed spectra, Rayleigh energies,
differential-equation residuals, and physical wavefunction constraints. No supervised
parameter-reconstruction term is included. 

\subsubsection{Total Objective}
\label{subsubsec:total_objective}

For a batch \(\mathcal{B}\) of nucleus--species systems, the complete
training objective is

\begin{equation}
\begin{aligned}
\mathcal{L}_{\mathrm{total}}
={}&
w_E\mathcal{L}_{E}
+
w_R\mathcal{L}_{R}
+
w_{\Theta}\mathcal{L}_{\Theta}
+
w_{\Phi}\mathcal{L}_{\Phi}
+
w_{\mathrm{BC}}\mathcal{L}_{\mathrm{BC}}
+
w_{\mathrm{orth}}\mathcal{L}_{\mathrm{orth}}
+
w_{\mathrm{SO}}\mathcal{L}_{\mathrm{SO}}
+
w_{\mathrm{KL}}\mathcal{L}_{\mathrm{KL}}.
\end{aligned}
\label{eq:total_loss}
\end{equation}

The physics losses are first averaged over the selected states of each
system and then over the systems in the batch. Unit normalization is
imposed by the differentiable scaling in
Eq.~\eqref{eq:normalization_factor}; therefore,
Eq.~\eqref{eq:total_loss} contains no normalization-loss term.

\subsection{Training and Inference}
\label{subsec:training_procedure}

WaveNet and ParamNet are optimized jointly. At every iteration, the
system-level ParamNet context is constructed using all selected
nucleus--species systems. ParamNet therefore constructs its global
variational distribution from the complete identification dataset, even
when a subset of systems is used to evaluate the physics-informed
objective.

A single latent vector is sampled and transformed into one physical
parameter vector. The same parameters are then used in every Hamiltonian
included in the current physics-loss batch. This distinction between the
full global context and the stochastic physics batch preserves global
parameter inference while reducing the cost of evaluating the
differential-equation losses.

The training procedure is summarized in
Algorithm~\ref{alg:qupi_pinn_training}.

\begin{algorithm}[H]
\caption{\textsc{QuPI-PINN}: Quantum Probabilistic Inverse PINN}
\label{alg:qupi_pinn_training}
\begin{algorithmic}[1]
\Require Global dataset
\(\mathcal{D}=\{\mathcal{D}_s\}_{s=1}^{N_{\mathrm{sys}}}\),
parameter bounds, quadrature grids, and number of iterations
\State Initialize WaveNet parameters \(\bm{\omega}\) and ParamNet
parameters \(\bm{\varphi}\)
\For{\(t=1,\ldots,N_{\mathrm{iter}}\)}
    \State Evaluate the separated WaveNet components for every selected
    state
    \State Compute the normalization scales and normalized radial
    functions
    \State Construct one ParamNet context vector for every system
    \State Encode all system contexts and calculate
    \(\bm{h}_{\mathrm{global}}\)
    \State Infer \(\bm{\mu}\) and \(\bm{\sigma}\)
    \State Sample
    \(\bm{z}=\bm{\mu}+\bm{\sigma}\odot\bm{\epsilon}\),
    \(\bm{\epsilon}\sim\mathcal{N}(\bm{0},\bm{I})\)
    \State Map \(\bm{z}\) to the bounded global parameter vector
    \(\bm{\vartheta}\)
    \State Select a system mini-batch \(\mathcal{B}\)
    \State Sample new radial and angular collocation points
    \State Evaluate
    \(\mathcal{L}_{E}\),
    \(\mathcal{L}_{R}\),
    \(\mathcal{L}_{\Theta}\),
    \(\mathcal{L}_{\Phi}\),
    \(\mathcal{L}_{\mathrm{BC}}\),
    \(\mathcal{L}_{\mathrm{orth}}\),
    \(\mathcal{L}_{\mathrm{SO}}\), and
    \(\mathcal{L}_{\mathrm{KL}}\)
    \State Form \(\mathcal{L}_{\mathrm{total}}\), backpropagate, and
    jointly update \(\bm{\omega}\) and \(\bm{\varphi}\)
\EndFor
\State \Return
\(\bm{\omega}^{*}\), \(\bm{\varphi}^{*}\)
\end{algorithmic}
\end{algorithm}

The two networks are optimized with Adam using separate learning rates.
Independent stepwise learning-rate schedulers are applied to WaveNet and
ParamNet. Gradients are clipped jointly according to

\begin{equation*}
\left\|
\nabla\mathcal{L}_{\mathrm{total}}
\right\|_2
\leq
5
\end{equation*}

to reduce instabilities associated with second derivatives and
stochastic latent sampling.

The quadrature grids used for wavefunction normalization and Rayleigh
energies are deterministic and fixed throughout training. By contrast,
the collocation points used for the differential residuals and the Monte
Carlo points used for orthogonality, are resampled at every iteration.
The network dimensions, learning rates, scheduler parameters, loss
weights, quadrature resolutions, and number of epochs are reported in
Table~\ref{tab:hpo_search_space}.

After training, the physical parameter distribution is evaluated through
Monte Carlo sampling. Latent samples are transformed using
Eq.~\eqref{eq:bounded_parameter_mapping}. The reported parameter
estimates are the sample means and standard deviations in physical
parameter space. These standard deviations quantify the spread of the
learned variational output distribution in physical parameter space.
They are controlled by the latent prior, KL weight, parameter bounds,
loss scaling, and optimization procedure, and are not interpreted as
calibrated Bayesian credible intervals for the Woods--Saxon parameters.
From a practical perspective, these standard deviations provide a
qualitative indication of parameter stiffness. They indicate which
Woods--Saxon parameters, such as the central depth, are tightly
constrained by the supplied single-particle spectra and which, such as
the surface diffuseness, lie in flatter loss basins and exhibit greater
functional flexibility.

Parameter-recovery accuracy in the synthetic experiments is quantified
using the absolute relative error

\begin{equation*}
\varepsilon_{\vartheta_k}
=
100
\frac{
\left|
\widehat{\vartheta}_k-\vartheta_k^{\mathrm{ref}}
\right|
}{
\left|\vartheta_k^{\mathrm{ref}}\right|
},
\end{equation*}

where \(\widehat{\vartheta}_k\) and
\(\vartheta_k^{\mathrm{ref}}\) denote the inferred and reference values
of parameter \(k\), respectively.

Spectral accuracy is evaluated from the state-level residual between the
calculated and target energies, where the target is either a synthetic
reference value or an experimental single-particle energy. For the
matched states, the mean absolute error (MAE), root-mean-square error (RMSE),
bias, and maximum absolute error \(\Delta E_{\max}\) are reported. The
bias retains the sign of the residual: negative values indicate that
the calculated states are, on average, more deeply bound than their
targets (underestimation), whereas positive values indicate less deeply
bound predictions (overestimation). All spectral metrics are computed
only over states matched according to the quantum numbers used for the
corresponding evaluation.


\FloatBarrier

\section{Results}
\label{sec:results}

All experiments used the six-parameter
Woods--Saxon vector defined in Eq.~\eqref{eq:six_parameter_vector} and
the common configuration summarized in
Table~\ref{tab:hpo_search_space}. The WaveNet and
ParamNet architectures, optimization settings, quadrature grids,
parameter bounds, and loss weights were selected by Bayesian
hyperparameter optimization and then held fixed across the synthetic
and experimental Seminole and Wahlborn experiments.


\subsection{Parameter Identification from Synthetic Spectra}
\label{subsec:synthetic_identification_results}

To evaluate the inverse-identification capability under controlled
conditions, synthetic single-particle spectra generated by the
independent finite-difference solver were supplied to the
physics-informed framework. The spectra were generated using the
ground-truth Seminole and Wahlborn parameter sets reported in
Table~\ref{tab:dataset_parameterizations}. For each functional form, the
objective was to recover the six generating Woods--Saxon parameters
from the corresponding synthetic spectra. The identification results
are evaluated first through direct parameter recovery and then through
a spectral closure test in which the inferred distribution-mean
parameters are reintroduced into the finite-difference solver.

\subsubsection{Parameter Recovery}
\label{subsubsec:synthetic_parameter_recovery}

The inferred physical-parameter distributions are compared with the
generating values in
Table~\ref{tab:synthetic_parameter_recovery}. The quoted
\(\sigma\) is the standard deviation obtained after sampling the learned
latent distribution and transforming the samples into physical
parameter space; it is reported as output spread rather than as a
calibrated credible interval.

\begin{table}[H]
\centering
\caption{
Recovery of the six global Woods--Saxon parameters in the Seminole and
Wahlborn synthetic experiments.
}
\label{tab:synthetic_parameter_recovery}
\scriptsize
\setlength{\tabcolsep}{3pt}
\renewcommand{\arraystretch}{1.08}
\begin{tabular}{lcccccccc}
\toprule
&
\multicolumn{4}{c}{Seminole}
&
\multicolumn{4}{c}{Wahlborn}
\\
\cmidrule(lr){2-5}
\cmidrule(lr){6-9}
Parameter
& Reference
& Distribution mean
& \(\sigma\)
& Rel. error [\%]
& Reference
& Distribution mean
& \(\sigma\)
& Rel. error [\%]
\\
\midrule
\(V_0\) [MeV]
& 52.06
& 52.09
& 0.0148
& 0.058
& 51.00
& 51.04
& 0.0131
& 0.078
\\
\(\kappa\)
& 0.639
& 0.640
& 0.0014
& 0.157
& 0.670
& 0.670
& 0.0013
& 0.000
\\
\(r_0\) [fm]
& 1.260
& 1.260
& 0.0004
& 0.000
& 1.270
& 1.270
& 0.0003
& 0.000
\\
\(a\) [fm]
& 0.662
& 0.665
& 0.0014
& 0.453
& 0.670
& 0.674
& 0.0012
& 0.597
\\
\(\lambda_{\mathrm{SO}}\)
& 24.1
& 24.1
& 0.1232
& 0.000
& 32.0
& 32.1
& 0.1335
& 0.313
\\
\(r_{0,\mathrm{SO}}\) [fm]
& 1.160
& 1.160
& 0.0023
& 0.000
& 1.270
& 1.271
& 0.0017
& 0.079
\\
\bottomrule
\end{tabular}
\tabnote{Distribution means are displayed at the precision shown;
relative errors are calculated from the unrounded numerical estimates.
The reported \(\sigma\) values quantify output spread rather than
calibrated credible intervals.}
\end{table}

All six Seminole parameters were recovered within \(0.5\%\), and all
Wahlborn parameters within \(0.6\%\). The largest deviation in both
experiments occurred for the surface diffuseness, with relative errors
of \(0.453\%\) and \(0.597\%\), respectively. Its slightly weaker
identification is consistent with the coupled role of \(a\) in the
central surface profile and, through \(a_{\mathrm{SO}}=a\), in the
spin--orbit interaction.

\subsubsection{Spectral Validation}
\label{subsubsec:synthetic_spectral_validation}

The recovered interactions were also assessed spectrally. For both
parameterizations, the distribution-mean parameters were reintroduced
into the independent finite-difference solver and compared state by
state with the reference interaction. The closure test covered bound
proton and neutron states in \(^{16}\mathrm{O}\),
\(^{40}\mathrm{Ca}\),
\(^{48}\mathrm{Ca}\), \(^{56}\mathrm{Ni}\),
\(^{100}\mathrm{Sn}\), \(^{132}\mathrm{Sn}\), and
\(^{208}\mathrm{Pb}\); the results are summarized in
Table~\ref{tab:synthetic_spectral_validation}.

\begin{table}[H]
\centering
\caption{
Finite-difference closure comparison for the Seminole and Wahlborn
parameterizations.
}
\label{tab:synthetic_spectral_validation}
\small
\setlength{\tabcolsep}{8pt}
\renewcommand{\arraystretch}{1.08}
\begin{tabular}{llcc}
\toprule
Parameterization
& Spectrum
& States
& MAE [MeV]
\\
\midrule
\multirow{3}{*}{Seminole}
& Neutrons
& 106
& 0.0096
\\
& Protons
& 81
& 0.0126
\\
& All states
& 187
& 0.0109
\\
\midrule
\multirow{3}{*}{Wahlborn}
& Neutrons
& 103
& 0.0135
\\
& Protons
& 78
& 0.0124
\\
& All states
& 181
& 0.0131
\\
\bottomrule
\end{tabular}
\end{table}

The closure MAE was \(0.0109~\mathrm{MeV}\) for Seminole and
\(0.0131~\mathrm{MeV}\) for Wahlborn, corresponding to average
state-level deviations of approximately \(11\) and \(13~\mathrm{keV}\).
Errors were similarly small in the neutron and proton subsets.

\subsection{Parameter Identification from Experimental Spectra}
\label{subsec:experimental_identification}

The inverse solver was next applied to experimental single-particle
spectra. In this case, the generating Woods--Saxon parameters are
unknown. The objective is therefore not to recover a prescribed
reference parameter vector but to infer a global parameter distribution
that provides an accurate spectral description within the selected
functional form. The terms Seminole and Wahlborn refer here to the
functional form of the interaction (potential). The reference parameter sets are
used only as spectral baselines and are not treated as
parameter-identification targets.


\subsubsection{Spectrum Reconstruction Accuracy}
\label{subsubsec:experimental_reconstruction}

The inferred global parameter distributions are reported in
Table~\ref{tab:experimental_inferred_parameters}. The reported
\(\sigma\) values denote the standard deviations of the sampled
physical parameters after the bounded nonlinear transformation.

\begin{table}[H]
\centering
\caption{
Global parameters inferred from the experimental spectra.
}
\label{tab:experimental_inferred_parameters}
\small
\setlength{\tabcolsep}{5pt}
\renewcommand{\arraystretch}{1.08}
\begin{tabular}{lcccc}
\toprule
&
\multicolumn{2}{c}{Seminole expression}
&
\multicolumn{2}{c}{Wahlborn expression}
\\
\cmidrule(lr){2-3}
\cmidrule(lr){4-5}
Parameter
& Distribution mean \(\pm\sigma\)
& Best sampled set
& Distribution mean \(\pm\sigma\)
& Best sampled set
\\
\midrule
\(V_0\) [MeV]
& \(54.150\pm0.014\)
& 54.143
& \(52.444\pm0.011\)
& 52.467
\\
\(\kappa\)
& \(0.612\pm0.001\)
& 0.611
& \(0.632\pm0.001\)
& 0.633
\\
\(r_0\) [fm]
& \(1.2300\pm0.0001\)
& 1.2299
& \(1.2633\pm0.0002\)
& 1.2601
\\
\(a\) [fm]
& \(0.6760\pm0.0007\)
& 0.6750
& \(0.6542\pm0.0009\)
& 0.6518
\\
\(\lambda_{\mathrm{SO}}\)
& \(25.27\pm0.07\)
& 25.35
& \(26.16\pm0.06\)
& 26.16
\\
\(r_{0,\mathrm{SO}}\) [fm]
& \(1.1360\pm0.0008\)
& 1.1373
& \(1.1901\pm0.0007\)
& 1.1907
\\
\bottomrule
\end{tabular}
\tabnote{Uncertainties denote standard deviations in physical parameter
space. The Seminole and Wahlborn reference values are omitted because
they are spectral baselines rather than parameter-identification
targets.}
\end{table}

Table~\ref{tab:experimental_spectral_metrics} compares the spectral
errors obtained from the reference parameter sets, the means of the
inferred distributions, and the best sampled parameter sets.

\begin{table}[H]
\centering
\caption{
Reconstruction accuracy against experimental single-particle energies.
}
\label{tab:experimental_spectral_metrics}
\scriptsize
\setlength{\tabcolsep}{5pt}
\renewcommand{\arraystretch}{1.08}
\begin{tabular}{lllcccc}
\toprule
Potential expression
& Evaluation
& Spectrum
& \(N\)
& MAE [MeV]
& RMSE [MeV]
& Bias [MeV]
\\
\midrule

\multirow{9}{*}{Seminole}
& \multirow{3}{*}{Reference values}
& All      & 94 & 0.7969 & 1.1025 & -0.2899 \\
&          & Neutrons & 59 & 0.8044 & 1.0708 & -0.2764 \\
&          & Protons  & 35 & 0.7843 & 1.1540 & -0.3126 \\
\cmidrule{2-7}

& \multirow{3}{*}{PINN distribution mean}
& All      & 94 & 0.8068 & 1.1530 & -0.3232 \\
&          & Neutrons & 59 & 0.8453 & 1.1492 & -0.3807 \\
&          & Protons  & 35 & 0.7420 & 1.1591 & -0.2263 \\
\cmidrule{2-7}

& \multirow{3}{*}{Best sampled set}
& All      & 94 & 0.8056 & 1.1525 & -0.3207 \\
&          & Neutrons & 59 & 0.8433 & 1.1486 & -0.3812 \\
&          & Protons  & 35 & 0.7420 & 1.1590 & -0.2186 \\
\midrule

\multirow{9}{*}{Wahlborn}
& \multirow{3}{*}{Reference values}
& All      & 94 & 1.0783 & 1.4113 & 0.8429 \\
&          & Neutrons & 59 & 1.0101 & 1.2944 & 0.7837 \\
&          & Protons  & 35 & 1.1933 & 1.5891 & 0.9426 \\
\cmidrule{2-7}

& \multirow{3}{*}{PINN distribution mean}
& All      & 94 & 0.8303 & 1.1036 & 0.2033 \\
&          & Neutrons & 59 & 0.7763 & 1.0297 & 0.0868 \\
&          & Protons  & 35 & 0.9228 & 1.2199 & 0.4030 \\
\cmidrule{2-7}

& \multirow{3}{*}{Best sampled set}
& All      & 94 & 0.8177 & 1.0840 & 0.1155 \\
&          & Neutrons & 59 & 0.7652 & 1.0183 & 0.0197 \\
&          & Protons  & 35 & 0.9076 & 1.1884 & 0.2799 \\
\bottomrule
\end{tabular}
\tabnote{The Seminole and Wahlborn reference parameter sets are used as
spectral baselines. \(N\) denotes the number of matched bound states.}
\end{table}

For the Wahlborn expression, parameter identification improves all
aggregate metrics relative to the reference set. The distribution mean
reduces the all-state MAE by \(23.0\%\), from \(1.0783\) to
\(0.8303~\mathrm{MeV}\), while the best sampled set gives a
\(24.2\%\) reduction, to \(0.8177~\mathrm{MeV}\). Because the MAE
measures the typical absolute spectral deviation, this indicates a
broad improvement across the matched states. The RMSE decreases by
\(21.8\%\) for the distribution mean and \(23.2\%\) for the best sampled
set, showing that the improvement also extends to the more strongly
mis-predicted levels, to which the RMSE is particularly sensitive. The
magnitude of the positive bias decreases by \(75.9\%\), from
\(0.8429\) to \(0.2033~\mathrm{MeV}\), and by \(86.3\%\), to
\(0.1155~\mathrm{MeV}\), for the best sampled set. Thus, the systematic
tendency of the Wahlborn reference set to predict energies that are too
high is strongly reduced. The neutron and proton MAEs decrease by
\(23.1\%\) and \(22.7\%\), respectively, for the distribution mean, and
by \(24.2\%\) and \(23.9\%\) for the best sampled set, confirming that
the improvement is not confined to one nucleon sector.

For the Seminole expression, the inferred parameters retain accuracy
close to the reference set but do not improve the aggregate metrics.
The all-state MAE increases by \(1.2\%\) for the distribution mean and
\(1.1\%\) for the best sampled set, while the RMSE increases by
approximately \(4.6\%\) in both cases. The larger relative change in
RMSE indicates that a limited number of states acquire somewhat larger
errors even though the average absolute deviation remains nearly
unchanged. The magnitude of the negative bias increases by \(11.5\%\)
for the distribution mean and \(10.6\%\) for the best sampled set,
indicating a slightly stronger overall tendency toward overbinding.
This behavior is sector dependent: the proton MAE improves by
\(5.4\%\), from \(0.7843\) to \(0.7420~\mathrm{MeV}\), whereas the
neutron MAE worsens by \(5.1\%\) for the distribution mean and
\(4.8\%\) for the best sampled set. The aggregate similarity therefore
results from improved proton predictions compensating for reduced
neutron accuracy.

Among the two inferred distribution means, Seminole gives the lower
all-state MAE by approximately \(2.8\%\), with
\(0.8068~\mathrm{MeV}\) compared with \(0.8303~\mathrm{MeV}\) for
Wahlborn. However, their RMSE values are much closer, differing by only
about \(4.3\%\). The main distinction is therefore that the Seminole
expression retains the better average spectral accuracy, whereas the
Wahlborn identification produces the larger improvement relative to
its own reference set and removes most of its original systematic bias.

\subsubsection{Spectral Performance Across Nuclei}
\label{subsubsec:spectral_performance_across_nuclei}

Table~\ref{tab:seminole_performance_by_nucleus} reports nucleus-resolved
errors for the Seminole reference and PINN-inferred Seminole
interactions. A dagger denotes a nucleus excluded from PINN training,
and a double dagger denotes a nucleus excluded from both the PINN
training set and the least-squares calibration of the Seminole
reference interaction.

\begin{table}[H]
\centering
\caption{
Spectral performance across nuclei for the Seminole reference and the
PINN-inferred Seminole interaction.
}
\label{tab:seminole_performance_by_nucleus}
\scriptsize
\setlength{\tabcolsep}{4pt}
\renewcommand{\arraystretch}{1.08}
\begin{tabular}{lccccccc}
\toprule
&
&
\multicolumn{3}{c}{Seminole reference}
&
\multicolumn{3}{c}{PINN-inferred Seminole}
\\
\cmidrule(lr){3-5}
\cmidrule(lr){6-8}
Nucleus or subset
& \(N\)
& MAE [MeV]
& RMSE [MeV]
& \(\Delta E_{\max}\) [MeV]
& MAE [MeV]
& RMSE [MeV]
& \(\Delta E_{\max}\) [MeV]
\\
\midrule
\(^{16}\mathrm{O}^{\dagger}\)
& 6 & 2.4937 & \textbf{2.7367} & \textbf{4.0663}
& \textbf{2.4421} & 2.7488 & 4.1140
\\
\(^{40}\mathrm{Ca}\)
& 11 & 1.0080 & 1.1499 & 2.1227
& \textbf{0.8061} & \textbf{1.0249} & \textbf{2.1153}
\\
\(^{48}\mathrm{Ca}\)
& 13 & \textbf{1.0431} & \textbf{1.2858} & \textbf{2.9647}
& 1.0981 & 1.4594 & 3.5882
\\
\(^{56}\mathrm{Ni}^{\dagger}\)
& 6 & 0.7782 & \textbf{0.9519} & \textbf{1.7235}
& \textbf{0.7744} & 0.9860 & 1.7669
\\
\(^{90}\mathrm{Zr}^{\dagger\ddagger}\)
& 10 & 0.8678 & \textbf{1.0010} & 1.8314
& \textbf{0.8407} & 1.0457 & \textbf{1.8173}
\\
\(^{100}\mathrm{Sn}^{\dagger}\)
& 11 & \textbf{0.7218} & \textbf{0.9793} & \textbf{2.5191}
& 0.8031 & 1.0678 & 2.7913
\\
\(^{132}\mathrm{Sn}\)
& 14 & \textbf{0.4157} & \textbf{0.4708} & \textbf{0.9636}
& 0.4431 & 0.5796 & 1.4184
\\
\(^{208}\mathrm{Pb}\)
& 23 & \textbf{0.3562} & \textbf{0.4256} & \textbf{0.9133}
& 0.4277 & 0.5217 & 1.3588
\\
\midrule
Neutrons
& 59 & \textbf{0.8044} & \textbf{1.0708} & \textbf{3.7952}
& 0.8433 & 1.1486 & 3.8900
\\
Protons
& 35 & 0.7843 & \textbf{1.1540} & \textbf{4.0663}
& \textbf{0.7420} & 1.1590 & 4.1140
\\
\midrule
All matched levels
& 94 & \textbf{0.7969} & \textbf{1.1025} & \textbf{4.0663}
& 0.8056 & 1.1525 & 4.1140
\\
\bottomrule
\end{tabular}
\tabnote{A dagger (\(\dagger\)) denotes a nucleus excluded from PINN
training; a double dagger (\(\ddagger\)) denotes a nucleus excluded from
both parameter-determination procedures. Bold indicates the smaller
error in each pairwise comparison.}
\end{table}

Across the complete \(94\)-state benchmark, the PINN-inferred Seminole
interaction gives an MAE of \(0.8056~\mathrm{MeV}\), compared with
\(0.7969~\mathrm{MeV}\) for the Seminole reference parameterization.
The difference is approximately \(1.1\%\) of the reference MAE. The
corresponding RMSE values are \(1.1525\) and \(1.1025~\mathrm{MeV}\),
while the maximum absolute errors differ by less than
\(0.05~\mathrm{MeV}\).

This comparable aggregate accuracy is obtained from a substantially
smaller identification dataset. The PINN uses \(42\) selected levels
from four nuclei, whereas the Seminole reference parameters were
determined from a global least-squares fit spanning seven nuclei. The
inferred interaction reduces the proton MAE from \(0.7843\) to
\(0.7420~\mathrm{MeV}\), while the neutron MAE increases from
\(0.8044\) to \(0.8433~\mathrm{MeV}\). At the nucleus level, the
inferred parameters produce lower MAEs for \(^{16}\mathrm{O}\),
\(^{40}\mathrm{Ca}\), \(^{56}\mathrm{Ni}\), and \(^{90}\mathrm{Zr}\),
whereas the reference parameters remain more accurate for the remaining
systems.

The \(^{90}\mathrm{Zr}\) comparison is the fully independent
out-of-calibration test because this nucleus was used by neither
parameter-determination procedure. The PINN-inferred interaction
reduces its MAE from \(0.8678\) to \(0.8407~\mathrm{MeV}\) and its
maximum absolute deviation from \(1.8314\) to
\(1.8173~\mathrm{MeV}\), although the Seminole reference
parameterization gives the lower RMSE.

Figure~\ref{fig:seminole_level_schemes} compares representative
experimental single-particle spectra with those obtained from the
Seminole reference parameters and the PINN best sampled set.

\noindent\begin{minipage}{\linewidth}
    \centering
    \captionsetup{type=figure}

    \begin{subfigure}[t]{0.47\linewidth}
        \centering
        \includegraphics[width=\linewidth]
        {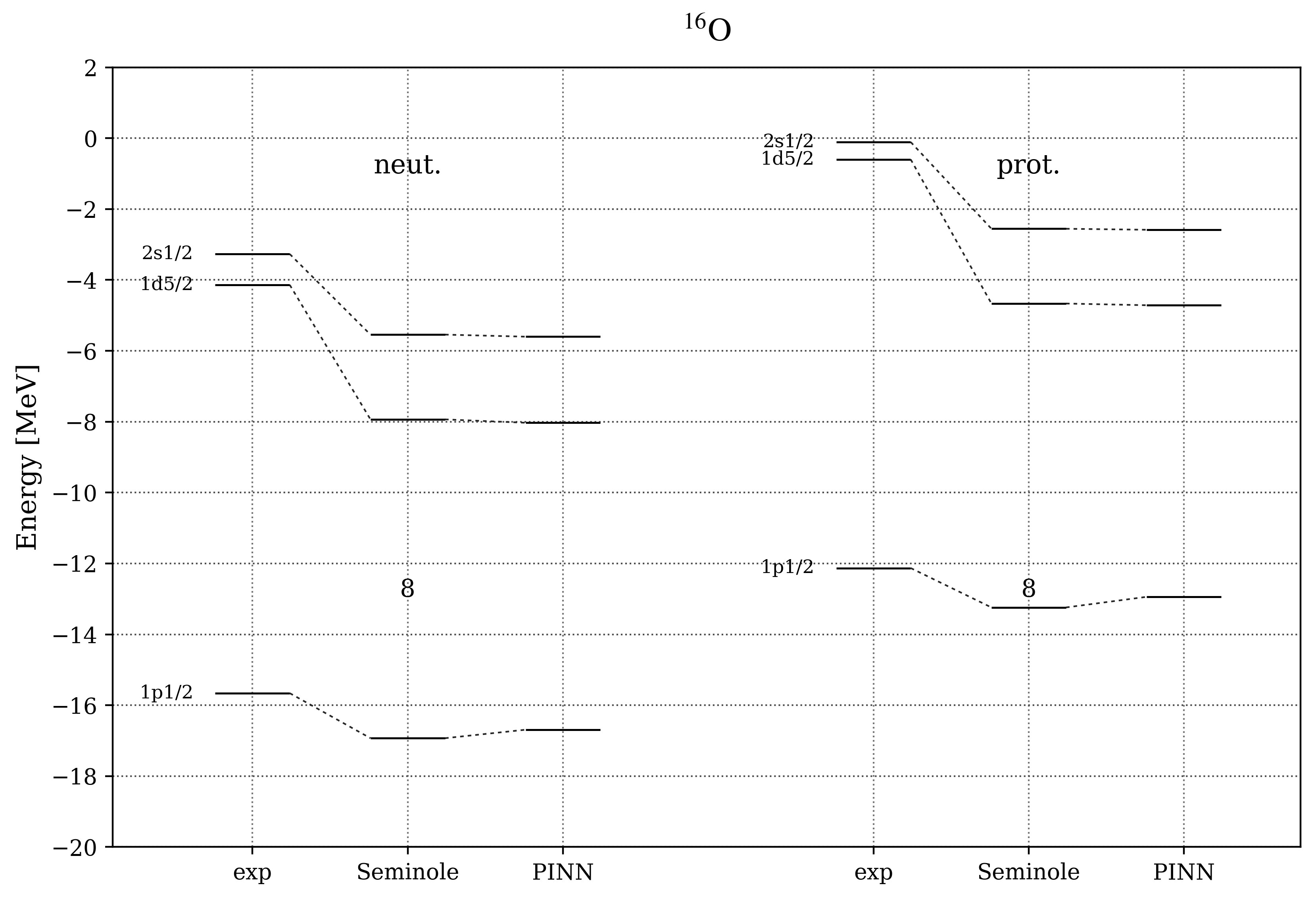}
        \caption{\(^{16}\mathrm{O}\)}
        \label{fig:levels_16O}
    \end{subfigure}
    \hfill
    \begin{subfigure}[t]{0.47\linewidth}
        \centering
        \includegraphics[width=\linewidth]
        {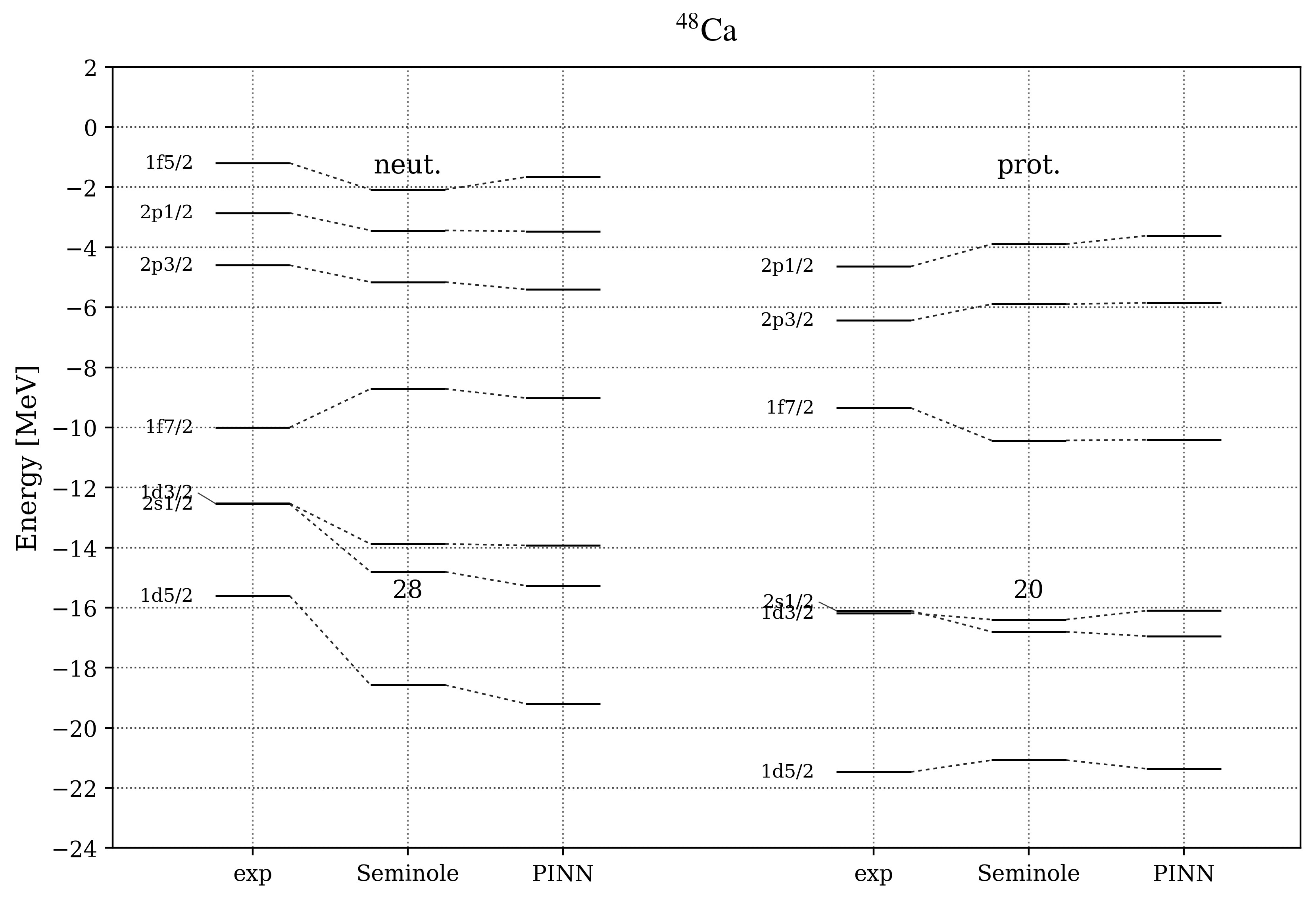}
        \caption{\(^{48}\mathrm{Ca}\)}
        \label{fig:levels_48Ca}
    \end{subfigure}

    \vspace{0.5em}

    \begin{subfigure}[t]{0.47\linewidth}
        \centering
        \includegraphics[width=\linewidth]
        {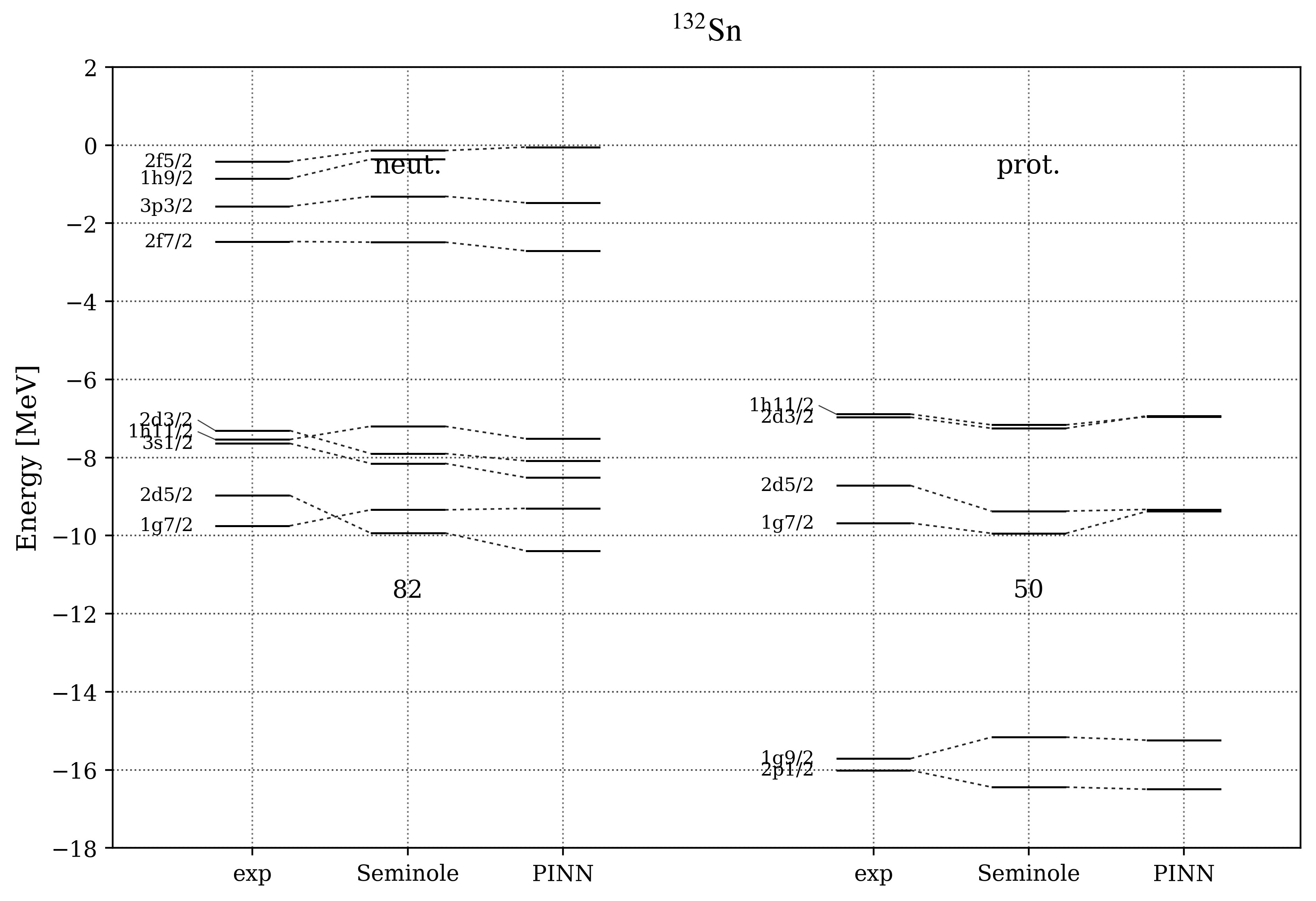}
        \caption{\(^{132}\mathrm{Sn}\)}
        \label{fig:levels_132Sn}
    \end{subfigure}
    \hfill
    \begin{subfigure}[t]{0.47\linewidth}
        \centering
        \includegraphics[width=\linewidth]
        {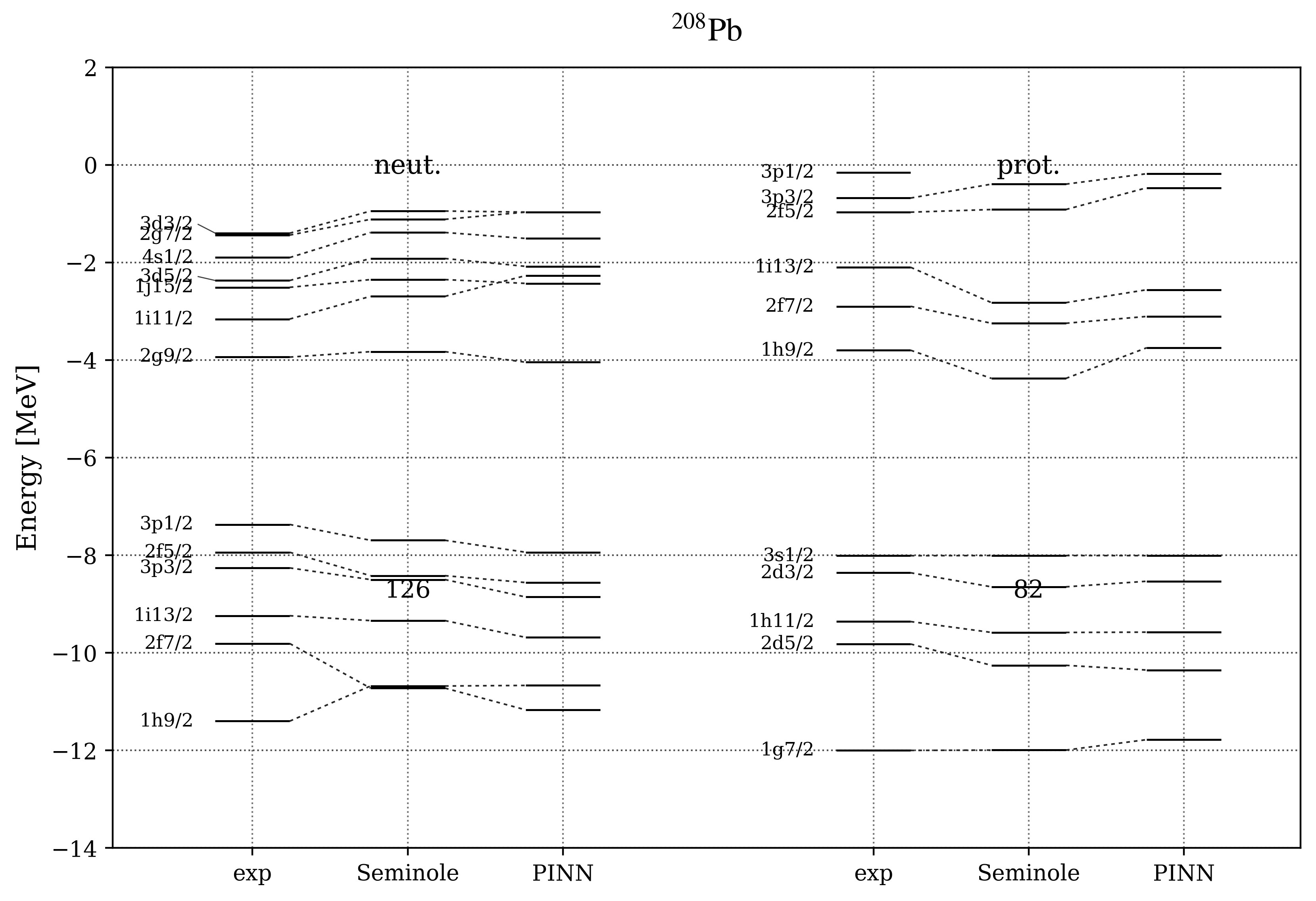}
        \caption{\(^{208}\mathrm{Pb}\)}
        \label{fig:levels_208Pb}
    \end{subfigure}

    \caption{
    Representative neutron and proton single-particle level schemes for
    (a) \(^{16}\mathrm{O}\), (b) \(^{48}\mathrm{Ca}\),
    (c) \(^{132}\mathrm{Sn}\), and (d) \(^{208}\mathrm{Pb}\), comparing
    experiment, the Seminole reference parameterization, and the PINN
    best sampled set. Horizontal segments indicate orbital energies,
    while dashed lines connect states with identical quantum-number
    assignments. The numbers between the neutron and proton spectra
    indicate the corresponding closed-shell particle numbers.
    }
    \label{fig:seminole_level_schemes}
\end{minipage}

\subsubsection{Spin--Orbit Splitting Consistency}
\label{subsubsec:spin_orbit_splitting_consistency}

The consistency of the inferred spin--orbit structure was examined by
comparing experimentally observed doublet splittings with those obtained
from the Seminole reference and the best-sample PINN interaction.
Spin--orbit doublets were identified as states belonging to the same
nucleus and nucleon species, with identical \((n_r,l)\) quantum numbers
and total angular momenta
\(j=l+\tfrac{1}{2}\) and \(j=l-\tfrac{1}{2}\). The splitting of each
matched doublet was defined as
\begin{equation}
\Delta E_{\mathrm{SO}}
=
E_{j=l+1/2}
-
E_{j=l-1/2}.
\label{eq:spin_orbit_splitting}
\end{equation}
The splitting error was subsequently evaluated
as
\begin{equation}
\epsilon_{\mathrm{SO}}
=
\Delta E_{\mathrm{SO}}^{\mathrm{calc}}
-
\Delta E_{\mathrm{SO}}^{\mathrm{exp}}.
\label{eq:spin_orbit_splitting_error}
\end{equation}

\begin{table}[H]
\centering
\caption{
Spin--orbit splitting errors relative to experiment for the Seminole
reference and best-sample PINN interactions.
}
\label{tab:spin_orbit_splitting_consistency}
\small
\setlength{\tabcolsep}{4.5pt}
\renewcommand{\arraystretch}{1.08}
\begin{tabular}{lrcccc}
\toprule
&
&
\multicolumn{2}{c}{Seminole}
&
\multicolumn{2}{c}{PINN}
\\
\cmidrule(lr){3-4}
\cmidrule(lr){5-6}

Subset
& \(N_{\mathrm d}\)
& MAE
& RMSE
& MAE
& RMSE
\\
\midrule
Neutron doublets
& 17
& \textbf{0.6423}
& \textbf{0.8671}
& 0.7331
& 0.9310
\\
Proton doublets
& 9
& 0.6433
& 0.7970
& \textbf{0.5217}
& \textbf{0.6305}
\\
\midrule
All doublets
& 26
& \textbf{0.6427}
& 0.8435
& 0.6599
& \textbf{0.8392}
\\
\bottomrule
\end{tabular}
\tabnote{\(N_{\mathrm d}\) denotes the number of matched doublets. Bold
indicates the smaller error. All error values are in MeV.}
\end{table}

As shown in Table~\ref{tab:spin_orbit_splitting_consistency}, the
Seminole reference reproduces neutron doublets more accurately, whereas
the PINN gives smaller errors for proton doublets. Over all \(26\)
doublets, the two interactions perform nearly identically: Seminole has
the slightly lower MAE, while the PINN has the slightly lower RMSE.
Thus, the PINN preserves the overall experimental spin--orbit splitting
structure with accuracy comparable to the Seminole reference.

\subsubsection{Comparison with Least-Squares Optimization}
\label{subsubsec:lsq_comparison}

The PINN-inferred interaction was compared with the least-squares fit
based on the finite-difference solver in
Sec.~\ref{subsec:fd_lsq_fit}. Both methods used the
same \(42\) identification levels, and their optimized parameter vectors
were evaluated with the same finite-difference solver on the \(94\)
common experimental states. Table~\ref{tab:pinn_lsq_comparison} reports
the nucleus-resolved MAE, RMSE, and maximum absolute deviation.

\begin{table}[H]
\centering
\caption{
Spectral errors for the finite-difference and PINN interactions, fitted
to the same \(42\) levels and evaluated on common bound states.
}
\label{tab:pinn_lsq_comparison}
\scriptsize
\setlength{\tabcolsep}{4.5pt}
\renewcommand{\arraystretch}{1.08}
\begin{tabular}{lrcccccc}
\toprule
&
&
\multicolumn{3}{c}{Finite-difference LSQ fit}
&
\multicolumn{3}{c}{PINN}
\\
\cmidrule(lr){3-5}
\cmidrule(lr){6-8}
Nucleus or subset
& \(N\)
& MAE
& RMSE
& \(\Delta E_{\max}\)
& MAE
& RMSE
& \(\Delta E_{\max}\)
\\
\midrule

\(^{16}\mathrm{O}^{\dagger}\)
& 6
& 2.6581
& 3.0811
& 4.7714
& \textbf{2.4421}
& \textbf{2.7488}
& \textbf{4.1140}
\\

\(^{40}\mathrm{Ca}\)
& 11
& 1.2642
& 1.4823
& 2.6602
& \textbf{0.8061}
& \textbf{1.0249}
& \textbf{2.1153}
\\

\(^{48}\mathrm{Ca}\)
& 13
& \textbf{0.7827}
& \textbf{0.9585}
& \textbf{1.8362}
& 1.0981
& 1.4594
& 3.5882
\\

\(^{56}\mathrm{Ni}^{\dagger}\)
& 6
& 0.8226
& 1.0376
& 1.9685
& \textbf{0.7744}
& \textbf{0.9860}
& \textbf{1.7669}
\\

\(^{90}\mathrm{Zr}^{\dagger\ddagger}\)
& 10
& \textbf{0.7852}
& \textbf{0.9567}
& 2.1578
& 0.8407
& 1.0457
& \textbf{1.8173}
\\

\(^{100}\mathrm{Sn}^{\dagger}\)
& 11
& \textbf{0.6733}
& \textbf{1.0079}
& \textbf{2.5785}
& 0.8031
& 1.0678
& 2.7913
\\

\(^{132}\mathrm{Sn}\)
& 14
& 0.4716
& \textbf{0.5642}
& \textbf{0.8918}
& \textbf{0.4431}
& 0.5796
& 1.4184
\\

\(^{208}\mathrm{Pb}\)
& 23
& \textbf{0.4054}
& 0.5411
& \textbf{1.2882}
& 0.4277
& \textbf{0.5217}
& 1.3588
\\

\midrule

Held-out reference nuclei
& 23
& 1.2300
& 1.8009
& 4.7714
& \textbf{1.2232}
& \textbf{1.6644}
& \textbf{4.1140}
\\

\midrule

Neutrons
& 59
& \textbf{0.7429}
& \textbf{1.0640}
& 4.3808
& 0.8433
& 1.1486
& \textbf{3.8900}
\\

Protons
& 35
& 0.9235
& 1.3549
& 4.7714
& \textbf{0.7420}
& \textbf{1.1590}
& \textbf{4.1140}
\\

\midrule

All common levels
& 94
& 0.8101
& 1.1807
& 4.7714
& \textbf{0.8056}
& \textbf{1.1525}
& \textbf{4.1140}
\\

\bottomrule
\end{tabular}
\tabnote{A dagger (\(\dagger\)) marks nuclei excluded from
identification; the double dagger (\(\ddagger\)) marks the independent
\(^{90}\mathrm{Zr}\) test. Bold indicates the smaller error. All error
values are in MeV.}
\end{table}

The two methods show closely comparable global accuracy. The PINN gives
an MAE and RMSE of \(0.8056\) and \(1.1525~\mathrm{MeV}\), respectively,
compared with \(0.8101\) and \(1.1807~\mathrm{MeV}\) for the
least-squares fit. For the \(23\) levels of the held-out reference
nuclei, their MAEs are nearly identical, while the lower PINN RMSE is
mainly associated with improved performance for \(^{16}\mathrm{O}\).
The nucleus-resolved results remain mixed, with neither method
consistently outperforming the other.

For the independent \(^{90}\mathrm{Zr}\) test, the least-squares fit
gives smaller MAE and RMSE, whereas the PINN gives a smaller maximum
deviation. Overall, finite-difference least-squares optimization provides
a strong conventional baseline, while the PINN retains comparable
spectral accuracy together with joint physics-constrained inference of
the interaction parameters and wavefunctions. 

The comparison should be interpreted in light of the different
estimators produced by the two procedures. The least-squares method
returns a single optimized parameter vector, whereas the probabilistic
PINN returns a distribution of admissible parameter vectors. Reporting
the best sampled PINN parameter set therefore evaluates the most accurate
spectral realization obtained from the learned distribution.

\subsubsection{Neutron--Proton Performance Pattern}
\label{subsubsec:species_performance}

Figure~\ref{fig:species_relative_changes} reveals a consistent
species-dependent pattern. Relative to the Seminole reference, the PINN
best sampled set reduces the proton MAE by \(5.4\%\), while the neutron
MAE and RMSE increase by \(4.8\%\) and \(7.3\%\), respectively; the
proton RMSE changes by only \(0.4\%\). Relative to the
finite-difference least-squares fit, the PINN reduces the proton MAE and
RMSE by \(19.7\%\) and \(14.5\%\), while the corresponding neutron
errors increase by \(13.5\%\) and \(8.0\%\). The possible origin of this
proton-favoring behavior is discussed in Section~\ref{sec:discussion}.

\noindent\begin{minipage}{\linewidth}
\centering
\captionsetup{type=figure}
\includegraphics[width=0.88\linewidth]
{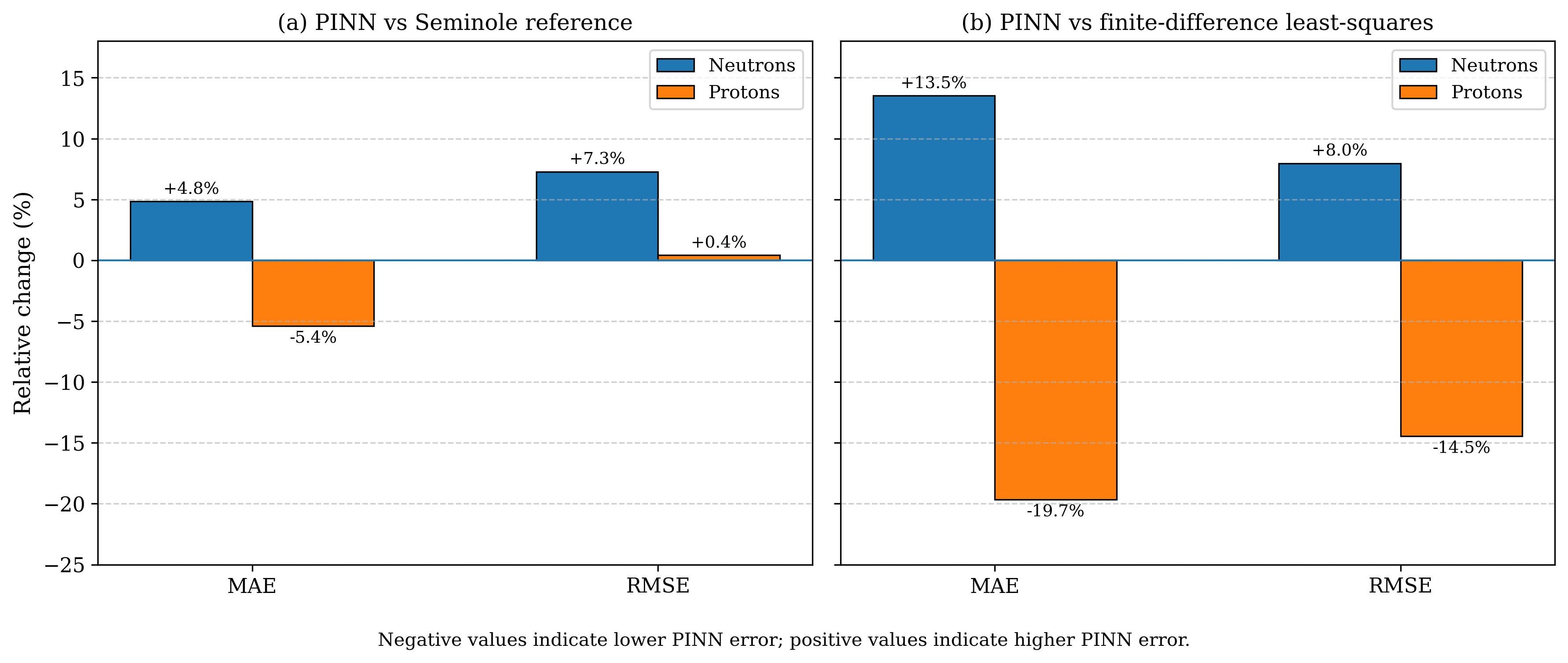}
\caption{
Species-resolved relative changes in the MAE and RMSE of the PINN
best sampled set with respect to (a) the Seminole reference and
(b) the finite-difference least-squares fit. Relative changes are
defined as
\(100(M_{\mathrm{PINN}}-M_{\mathrm{baseline}})
/M_{\mathrm{baseline}}\). 
}
\label{fig:species_relative_changes}
\end{minipage}
 
\subsection{Forward Solution of the Woods--Saxon Eigenvalue Problem}
\label{subsec:forward_results}

As a complementary diagnostic, the forward solution was evaluated
through the wavefunctions predicted by WaveNet for representative
proton states of \(^{208}\mathrm{Pb}\).

\noindent\begin{minipage}{\linewidth}
\centering
\captionsetup{type=figure}

\begin{subfigure}[t]{0.42\linewidth}
    \centering
    \includegraphics[width=\linewidth]
    {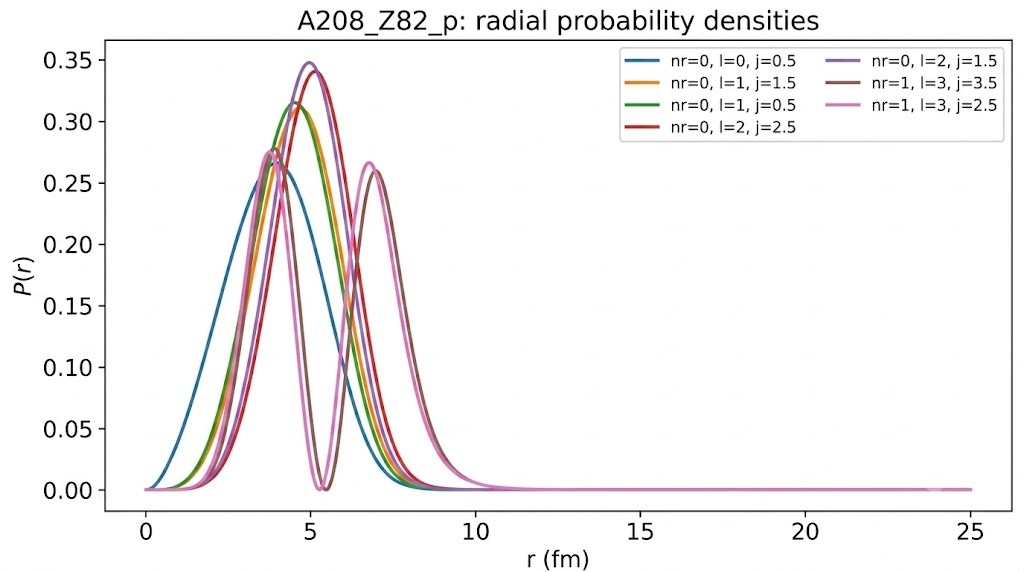}
    \caption{
    Radial probability densities \(P(r)\) for selected proton states.
    }
    \label{fig:pb208_proton_radial}
\end{subfigure}
\hfill
\begin{subfigure}[t]{0.42\linewidth}
    \centering
    \includegraphics[width=\linewidth]
    {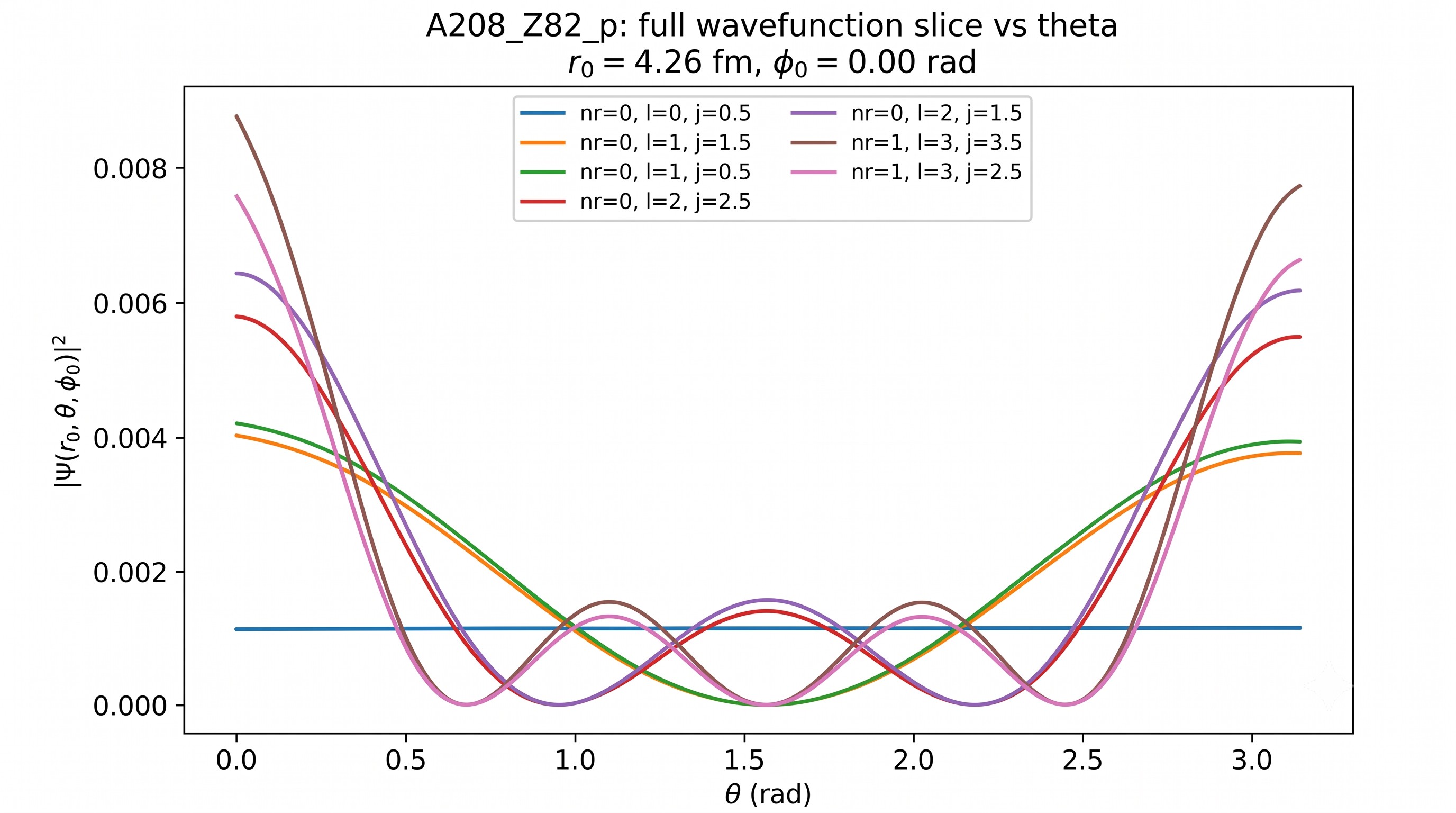}
    \caption{
    Total probability-density slices
    \(\lvert\Psi(r_0,\theta,\phi_0)\rvert^2\) at
    \(r_0=4.26~\mathrm{fm}\) and \(\phi_0=0\).
    }
    \label{fig:pb208_proton_theta}
\end{subfigure}

\vspace{0.3em}

\begin{subfigure}[t]{0.32\linewidth}
    \centering
    \includegraphics[width=\linewidth]
    {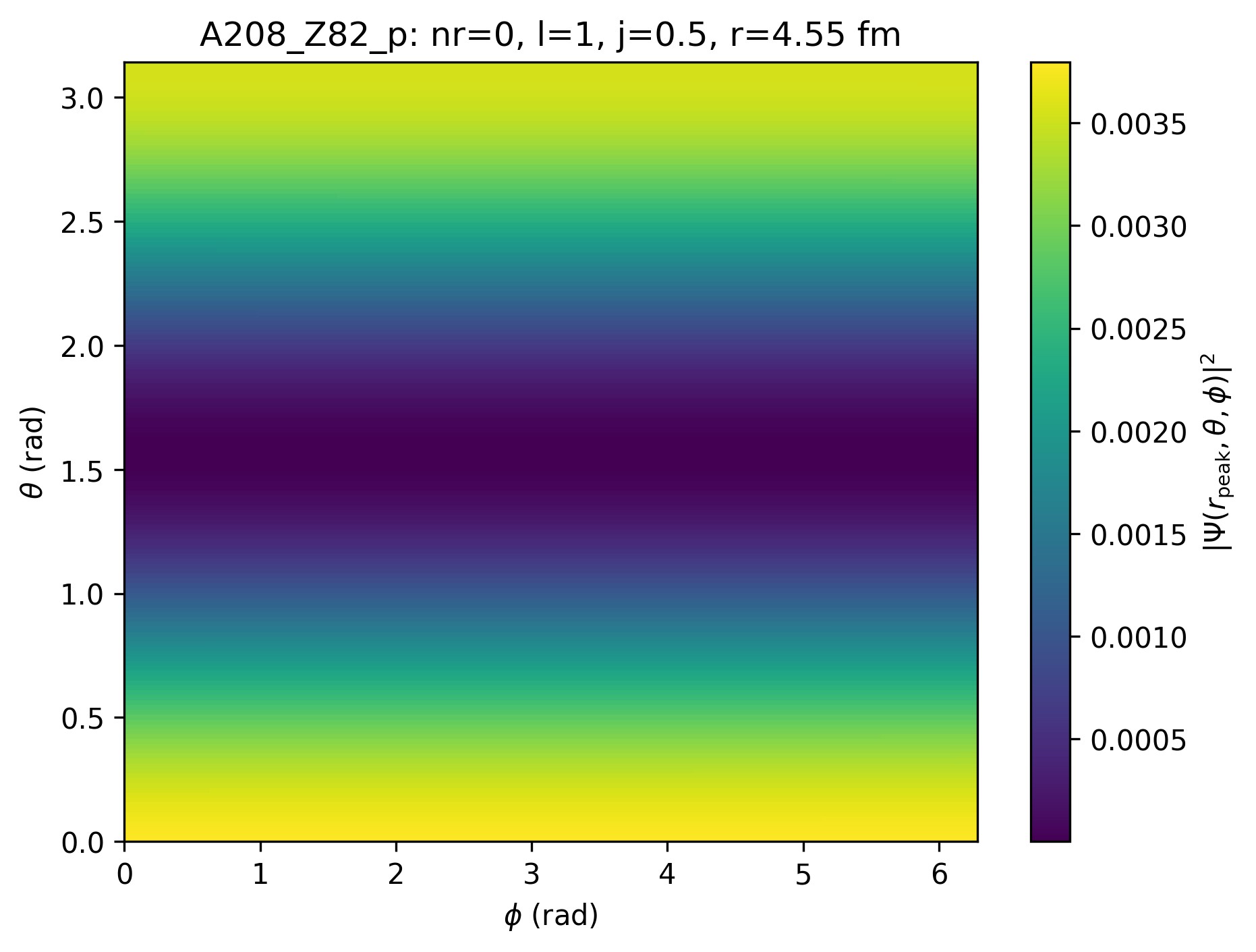}
    \caption{
    Total probability density of the
    \(1p_{1/2}\) state at \(r_{\mathrm{peak}}=4.55~\mathrm{fm}\).
    }
    \label{fig:pb208_proton_heatmap}
\end{subfigure}

\caption{
WaveNet forward-solution diagnostics for representative proton states
of \(^{208}\mathrm{Pb}\): (a) radial probability densities integrated
over the angular coordinates, (b) total probability-density slices as
functions of \(\theta\) at fixed \(r_0\) and \(\phi_0\), and
(c) the total probability density of the \(1p_{1/2}\) state over the
\((\theta,\phi)\) domain at fixed radius.
}
\label{fig:pb208_proton_forward}
\end{minipage}
Figure~\ref{fig:pb208_proton_forward} presents the radial probability
densities and two fixed-coordinate representations of the total spatial
probability density. As shown in Figure~\ref{fig:pb208_proton_radial}, the learned radial
profiles exhibit the expected dependence on the radial and orbital
quantum numbers. The \(1s_{1/2}\) state has no interior radial node,
whereas the \(2f_{7/2}\) and \(2f_{5/2}\) states each contain one node,
consistent with \(n_r=1\). Increasing \(l\) shifts the dominant
probability toward larger radii because of the centrifugal barrier.
Members of the same spin--orbit doublet have similar radial nodal
structures, with small differences arising from their distinct
\(\langle\bm{l}\cdot\bm{s}\rangle\) eigenvalues.

Figure~\ref{fig:pb208_proton_theta} shows slices of the total probability
density $\left|\Psi(r_0,\theta,\phi_0)\right|^2$ evaluated at the common coordinates
\(r_0=4.26~\mathrm{fm}\) and \(\phi_0=0\). The \(s\)-state density is
approximately independent of \(\theta\), whereas the \(p\)-, \(d\)-,
and \(f\)-state densities exhibit one, two, and three angular nodes,
respectively. Spin--orbit partners share the same angular nodal pattern
because they have the same \(l\), while their amplitudes differ because
of their state-dependent radial components, as expected from theoretical analysis.

Figure~\ref{fig:pb208_proton_heatmap} displays the total probability
density for the \(1p_{1/2}\) state at its radial-density maximum,
\(r_{\mathrm{peak}}=4.55~\mathrm{fm}\). For the selected \(m_l=0\)
component, the density is independent of \(\phi\) and vanishes at
\(\theta=\pi/2\), consistently with the expected \(l=1\) spatial
structure.

Overall, the probabilistic physics-informed solver retains aggregate
spectral accuracy comparable to both the finite-difference
least-squares fit and the Seminole reference while using a reduced set
of selected levels. The comparison reflects the different estimators:
least squares returns one optimized parameter vector, whereas the PINN
returns a learned distribution, from which the most accurate sampled
set is reported. The least-squares fit remains a strong baseline,
while the PINN additionally infers wavefunction representations under
the imposed physical constraints. The Wahlborn results further show
that the method can reduce systematic spectral shifts when the reference
set is not optimized for the selected benchmark.



\FloatBarrier

\section{Discussion}
\label{sec:discussion}

The results show that the proposed probabilistic physics-informed
framework can infer global Woods--Saxon parameters from bound
single-particle spectra while preserving the principal constraints of
the Schrödinger eigenvalue problem. In the synthetic closure tests, all
six parameters are recovered with sub-percent relative errors, and the
distribution-mean parameters reproduce the reference spectra in an
independent finite-difference solver with mean absolute deviations of
\(0.0109~\mathrm{MeV}\) for Seminole and \(0.0131~\mathrm{MeV}\) for
Wahlborn. These results confirm that the inferred parameter sets define
consistent Hamiltonians independently of the neural Rayleigh-energy
calculation.

For the experimental spectra, the framework does not establish a
substantial aggregate accuracy advantage over established methods. The
inferred Seminole interaction remains close to the Seminole reference,
while the finite-difference least-squares fit provides a similarly
accurate and computationally efficient baseline. In contrast, the
approximately \(23\%\) reduction of the Wahlborn all-state MAE shows
that the framework can substantially adapt a less accurate reference
expression to the selected spectral data. This improvement should be
interpreted as evidence of flexibility within the chosen functional
form rather than as general superiority over classical optimization.

A consistent species-dependent trend is observed, with the PINN giving
lower proton-sector errors but larger average neutron-sector errors.
One possible explanation is the additional Coulomb contribution in the
proton Hamiltonian. This term provides an extra physics-informed
constraint during training, which may help the network distinguish the
shared potential parameters more effectively and improve the proton
fit. The additional Coulomb dependence may therefore increase the
sensitivity of the proton spectra to the inferred parameters. This
interpretation remains qualitative, since the corresponding
species-resolved loss and gradient contributions were not analyzed
directly.

The main strength of the framework lies in its formulation of the
inverse problem. Parameter identification and the forward eigenvalue
problem are treated within one differentiable model: the global
interaction is inferred from spectral observations while the associated
energies and wavefunctions are constrained by the governing equation,
boundary conditions, normalization, orthogonality, and spin--orbit
structure. The outputs are therefore physically interpretable
interaction parameters and wavefunction representations rather than
direct, unconstrained energy predictions. Reintroducing the inferred
parameters into an independent finite-difference Hamiltonian provides a
separate consistency check.

The probabilistic output provides an additional distinction from a
deterministic fit. ParamNet returns a learned distribution over the
physical parameters, allowing multiple admissible parameter sets to be
sampled around the inferred solution. The associated standard
deviations indicate how concentrated the model output is around each
parameter value under the supplied spectra and physical constraints.
Smaller spreads suggest stronger model-derived confidence and greater
qualitative parameter stiffness, whereas larger spreads indicate
flatter effective loss directions and greater admissible flexibility.
These quantities are not calibrated Bayesian credible intervals, and
their interpretation remains conditional on the architecture, loss
weights, parameter bounds, latent prior, and optimization procedure.

A further strength is the feasibility of identification from limited
spectral information. The experimental study uses \(42\) selected
levels, compared with approximately \(86\) adopted levels in the
Seminole reference calibration, while retaining comparable aggregate
accuracy. Because the datasets and fitting procedures are not
identical, this comparison does not constitute controlled proof that the
method universally requires \(51\%\) fewer data. It nevertheless shows
that a structured subset of levels spanning different nuclei, nucleon
species, binding regimes, and spin--orbit doublets can constrain a
global interaction without nucleus-specific refitting. In addition, because the inverse
solver is implemented as a trainable neural architecture, the learned
WaveNet and ParamNet weights can be reused as an initialization and
fine-tuned when new spectral data become available, potentially reducing
the cost of subsequent identification tasks.

The main limitations arise from the small number of suitable
experimental single-particle datasets and from the limited analysis of
the observed neutron--proton performance difference. Because only a few
independent datasets were available, the robustness of the inferred
parameters across different experimental samples could not be examined
systematically. In addition, although the PINN performs better for
proton states in both comparisons, the reason for this behavior was not
studied directly. The possible influence of the Coulomb term therefore
remains a qualitative explanation that should be tested in future work.

Future work should test the framework with larger and noisier datasets,
repeated data partitions, sector-resolved gradient analyses, and
additional conventional baselines. Extensions to deformed, nonlocal,
energy-dependent, and continuum Hamiltonians would broaden the present
spherical bound-state formulation. The same forward--inverse strategy
could also be applied to other quantum problems involving potential
identification from sparse spectra, including molecular vibrations,
quantum wells, anharmonic oscillators, coupled channels, and inverse
scattering.

\section{Conclusion}

This study demonstrates that a probabilistic physics-informed neural
framework can recover a global Woods--Saxon parameterization from bound
single-particle spectra while simultaneously constructing
wavefunctions and energies that satisfy the principal constraints of
the Schrödinger eigenvalue problem. Controlled closure tests show that
the inverse procedure can identify the underlying interaction, and
independent finite-difference calculations confirm that the inferred
parameters remain physically consistent outside the neural solver. Application to experimental spectra shows that a limited but
physically structured collection of levels can constrain a global
interaction with spectral accuracy comparable to established
parameterizations and conventional least-squares optimization. The method's value instead lies in treating parameter
identification, forward eigenvalue solution, physical regularization,
and parameter variability within a unified differentiable model. The
inferred interaction can also be transferred to nuclei excluded from
identification without nucleus-specific refitting, indicating that the
learned result represents a global physical parameterization rather
than a collection of local spectral corrections. The probabilistic output provides a quantitative description of which
parameter directions are strongly or weakly constrained by the
available spectra. Nevertheless,
the results establish a alternative route to conventional methods,
particularly for inverse quantum problems in which observations are
sparse and physically admissible wavefunctions are required alongside
the inferred Hamiltonian parameters.

\FloatBarrier

\section*{Declaration of generative AI and AI-assisted technologies in the manuscript preparation process}

During the preparation of this manuscript, the authors used ChatGPT to
improve clarity and language consistency and to assist with code
generation. ChatGPT was not used to formulate the original scientific
ideas, research questions, methodology, or interpretation of the
results presented in this work. After using this tool, the authors
reviewed and verified the generated text and code and take full
responsibility for the integrity, accuracy, and reproducibility of the
final manuscript.

\printcredits

\section*{Declaration of Competing Interest}

The authors declare that they have no known competing financial
interests or personal relationships that could have appeared to
influence the work reported in this paper.

\section*{Acknowledgments}

This research did not receive any specific grant from funding agencies in the public, commercial, or not-for-profit sectors.

\FloatBarrier

\appendix

\section{Synthetic Parameter-Identification Dataset}
\label{app:synthetic_states}

The states used in the synthetic Seminole and Wahlborn
parameter-identification experiments are reported in
Tables~\ref{tab:seminole_selected_neutron_states},
\ref{tab:wahlborn_proton_states}, and
\ref{tab:wahlborn_neutron_states}.

\begin{table}[H]
\centering
\caption{
Neutron states used in the synthetic Seminole
parameter-identification experiment.
}
\label{tab:seminole_selected_neutron_states}
\scriptsize
\setlength{\tabcolsep}{5pt}
\renewcommand{\arraystretch}{1.08}
\begin{tabular}{cccccccc}
\toprule
\multicolumn{2}{c}{\(^{208}\mathrm{Pb}\)}
&
\multicolumn{2}{c}{\(^{132}\mathrm{Sn}\)}
&
\multicolumn{2}{c}{\(^{40}\mathrm{Ca}\)}
&
\multicolumn{2}{c}{\(^{48}\mathrm{Ca}\)}
\\
\cmidrule(lr){1-2}
\cmidrule(lr){3-4}
\cmidrule(lr){5-6}
\cmidrule(lr){7-8}
State & \(E^{\mathrm{FD}}\) [MeV]
&
State & \(E^{\mathrm{FD}}\) [MeV]
&
State & \(E^{\mathrm{FD}}\) [MeV]
&
State & \(E^{\mathrm{FD}}\) [MeV]
\\
\midrule
\(1s_{1/2}\) & -41.0393 &
\(1s_{1/2}\) & -38.6590 &
\(1s_{1/2}\) & -41.7902 &
\(1s_{1/2}\) & -36.3535
\\
\(1p_{3/2}\) & -37.1982 &
\(1p_{3/2}\) & -33.6866 &
\(1p_{3/2}\) & -32.0092 &
\(1p_{3/2}\) & -27.8574
\\
\(1p_{1/2}\) & -36.5966 &
\(1p_{1/2}\) & -32.7193 &
\(1p_{1/2}\) & -29.2469 &
\(1p_{1/2}\) & -25.4261
\\
\(1d_{5/2}\) & -32.6967 &
\(1d_{5/2}\) & -27.9673 &
\(1d_{5/2}\) & -21.3896 &
\(1d_{5/2}\) & -18.5747
\\
\(1d_{3/2}\) & -31.2875 &
\(1d_{3/2}\) & -25.8019 &
\(1d_{3/2}\) & -16.1485 &
\(1d_{3/2}\) & -13.8788
\\
\(3d_{5/2}\) & -1.9197 &
\(3p_{3/2}\) & -1.3125 &
\(2p_{3/2}\) & -6.4763 &
\(2p_{3/2}\) & -5.1609
\\
\(3d_{3/2}\) & -0.9470 &
\(3p_{1/2}\) & -0.7038 &
\(2p_{1/2}\) & -4.5101 &
\(2p_{1/2}\) & -3.4418
\\
\bottomrule
\end{tabular}
\end{table}

\begin{table}[H]
\centering
\caption{
Proton states used in the synthetic Wahlborn
parameter-identification experiment.
}
\label{tab:wahlborn_proton_states}
\scriptsize
\setlength{\tabcolsep}{5pt}
\renewcommand{\arraystretch}{1.08}
\begin{tabular}{cccccccc}
\toprule
\multicolumn{2}{c}{\(^{208}\mathrm{Pb}\)}
&
\multicolumn{2}{c}{\(^{132}\mathrm{Sn}\)}
&
\multicolumn{2}{c}{\(^{40}\mathrm{Ca}\)}
&
\multicolumn{2}{c}{\(^{48}\mathrm{Ca}\)}
\\
\cmidrule(lr){1-2}
\cmidrule(lr){3-4}
\cmidrule(lr){5-6}
\cmidrule(lr){7-8}
State & \(E^{\mathrm{FD}}\) [MeV]
&
State & \(E^{\mathrm{FD}}\) [MeV]
&
State & \(E^{\mathrm{FD}}\) [MeV]
&
State & \(E^{\mathrm{FD}}\) [MeV]
\\
\midrule
\(1s_{1/2}\) & -32.8158 &
\(1s_{1/2}\) & -38.4547 &
\(1s_{1/2}\) & -30.2343 &
\(1s_{1/2}\) & -37.2376
\\
\(1p_{3/2}\) & -29.5129 &
\(1p_{3/2}\) & -33.6765 &
\(1p_{3/2}\) & -21.3723 &
\(1p_{3/2}\) & -28.6838
\\
\(1p_{1/2}\) & -28.9733 &
\(1p_{1/2}\) & -32.8098 &
\(1p_{1/2}\) & -18.5460 &
\(1p_{1/2}\) & -26.1502
\\
\(1d_{5/2}\) & -25.3283 &
\(1d_{5/2}\) & -27.9737 &
\(1d_{5/2}\) & -11.8703 &
\(1d_{5/2}\) & -19.3581
\\
\(1d_{3/2}\) & -24.0737 &
\(1d_{3/2}\) & -25.9610 &
\(2s_{1/2}\) & -7.8035 &
\(1d_{3/2}\) & -14.0346
\\
\(2f_{7/2}\) & -3.0079 &
\(2d_{5/2}\) & -8.8834 &
\(1d_{3/2}\) & -6.2118 &
\(2p_{3/2}\) & -4.7642
\\
\(2f_{5/2}\) & -0.0468 &
\(2d_{3/2}\) & -6.1285 &
\(1f_{7/2}\) & -2.0384 &
\(2p_{1/2}\) & -2.0595
\\
\bottomrule
\end{tabular}
\end{table}

\begin{table}[H]
\centering
\caption{
Neutron states used in the synthetic Wahlborn
parameter-identification experiment.
}
\label{tab:wahlborn_neutron_states}
\scriptsize
\setlength{\tabcolsep}{5pt}
\renewcommand{\arraystretch}{1.08}
\begin{tabular}{cccccccc}
\toprule
\multicolumn{2}{c}{\(^{208}\mathrm{Pb}\)}
&
\multicolumn{2}{c}{\(^{132}\mathrm{Sn}\)}
&
\multicolumn{2}{c}{\(^{40}\mathrm{Ca}\)}
&
\multicolumn{2}{c}{\(^{48}\mathrm{Ca}\)}
\\
\cmidrule(lr){1-2}
\cmidrule(lr){3-4}
\cmidrule(lr){5-6}
\cmidrule(lr){7-8}
State & \(E^{\mathrm{FD}}\) [MeV]
&
State & \(E^{\mathrm{FD}}\) [MeV]
&
State & \(E^{\mathrm{FD}}\) [MeV]
&
State & \(E^{\mathrm{FD}}\) [MeV]
\\
\midrule
\(1s_{1/2}\) & -39.6998 &
\(1s_{1/2}\) & -37.2320 &
\(1s_{1/2}\) & -38.6437 &
\(1s_{1/2}\) & -34.6931
\\
\(1p_{3/2}\) & -35.8514 &
\(1p_{3/2}\) & -32.2547 &
\(1p_{3/2}\) & -29.2270 &
\(1p_{3/2}\) & -26.3142
\\
\(1p_{1/2}\) & -35.4823 &
\(1p_{1/2}\) & -31.6074 &
\(1p_{1/2}\) & -26.4896 &
\(1p_{1/2}\) & -24.2187
\\
\(1d_{5/2}\) & -31.3162 &
\(1d_{5/2}\) & -26.5204 &
\(1d_{5/2}\) & -19.1977 &
\(1d_{5/2}\) & -17.2871
\\
\(1d_{3/2}\) & -30.4028 &
\(1d_{3/2}\) & -24.9791 &
\(1d_{3/2}\) & -13.6266 &
\(1d_{3/2}\) & -12.9240
\\
\(3d_{5/2}\) & -1.6348 &
\(3p_{3/2}\) & -1.0017 &
\(2p_{3/2}\) & -5.1548 &
\(2p_{3/2}\) & -4.5615
\\
\(3d_{3/2}\) & -0.4426 &
\(3p_{1/2}\) & -0.3156 &
\(2p_{1/2}\) & -2.7707 &
\(2p_{1/2}\) & -2.6358
\\
\bottomrule
\end{tabular}
\end{table}

\subsection{Experimental Identification Dataset}
\label{app:experimental_identification_dataset}

Table~\ref{tab:complete_experimental_dataset} lists the 
\(96\) experimental single-particle energies used for training and evaluation purposes.
The dataset contains six states from each of seven nucleus--species
systems.

\begin{table*}[p]
\centering
\caption{
Complete experimental single-particle dataset used for parameter
identification and spectral evaluation. Energies are in MeV.
}
\label{tab:complete_experimental_dataset}

\fontsize{6.5}{6.2}\selectfont
\setlength{\tabcolsep}{1.8pt}
\renewcommand{\arraystretch}{0.80}

\begin{minipage}[t]{0.485\textwidth}
\centering
\begin{tabular}{clcrc}
\toprule
Nucleus & Sp. & Orbital & \(E^{\mathrm{obs}}\) & Set \\
\midrule

\(^{16}\mathrm{O}\) & \(n\) & \(1p_{1/2}\) & -15.66 & E \\
 &  & \(1d_{5/2}\) & -4.14 & E \\
 &  & \(2s_{1/2}\) & -3.27 & E \\
 & \(p\) & \(1p_{1/2}\) & -12.13 & E \\
 &  & \(1d_{5/2}\) & -0.60 & E \\
 &  & \(2s_{1/2}\) & -0.11 & E \\
\midrule

\(^{40}\mathrm{Ca}\) & \(n\) & \(1d_{5/2}\) & -22.39 & ID \\
 &  & \(2s_{1/2}\) & -18.19 & ID \\
 &  & \(1d_{3/2}\) & -15.64 & ID \\
 &  & \(1f_{7/2}\) & -8.36 & ID \\
 &  & \(2p_{3/2}\) & -5.84 & ID \\
 &  & \(2p_{1/2}\) & -4.20 & E \\
 &  & \(1f_{5/2}\) & -1.56 & ID \\
 & \(p\) & \(1d_{5/2}\) & -15.07 & E \\
 &  & \(2s_{1/2}\) & -10.92 & E \\
 &  & \(1d_{3/2}\) & -8.33 & E \\
 &  & \(1f_{7/2}\) & -1.09 & E \\
\midrule

\(^{48}\mathrm{Ca}\) & \(n\) & \(1d_{5/2}\) & -15.61 & ID \\
 &  & \(2s_{1/2}\) & -12.55 & ID \\
 &  & \(1d_{3/2}\) & -12.53 & ID \\
 &  & \(1f_{7/2}\) & -10.00 & ID \\
 &  & \(2p_{3/2}\) & -4.60 & ID \\
 &  & \(2p_{1/2}\) & -2.86 & ID \\
 &  & \(1f_{5/2}\) & -1.20 & E \\
 & \(p\) & \(1d_{5/2}\) & -21.47 & ID \\
 &  & \(1d_{3/2}\) & -16.18 & ID \\
 &  & \(2s_{1/2}\) & -16.10 & ID \\
 &  & \(1f_{7/2}\) & -9.35 & ID \\
 &  & \(2p_{3/2}\) & -6.44 & ID \\
 &  & \(2p_{1/2}\) & -4.64 & ID \\
\midrule

\(^{56}\mathrm{Ni}\) & \(n\) & \(1f_{7/2}\) & -16.64 & E \\
 &  & \(2p_{3/2}\) & -10.25 & E \\
 &  & \(1f_{5/2}\) & -9.48 & E \\
 &  & \(2p_{1/2}\) & -9.13 & E \\
 & \(p\) & \(1f_{7/2}\) & -7.17 & E \\
 &  & \(2p_{3/2}\) & -0.69 & E \\
\midrule

\(^{90}\mathrm{Zr}\) & \(n\) & \(1f_{7/2}\) & -21.00 & E \\
 &  & \(2p_{3/2}\) & -14.50 & E \\
 &  & \(1f_{5/2}\) & -14.10 & E \\
 &  & \(2p_{1/2}\) & -12.60 & E \\
 &  & \(1g_{9/2}\) & -11.97 & E \\
 &  & \(2d_{5/2}\) & -7.19 & E \\
 &  & \(3s_{1/2}\) & -6.00 & E \\
 &  & \(2d_{3/2}\) & -5.10 & E \\
 &  & \(1g_{7/2}\) & -4.80 & E \\
 &  & \(1h_{11/2}\) & -4.40 & E \\

\bottomrule
\end{tabular}
\end{minipage}
\hfill
\begin{minipage}[t]{0.485\textwidth}
\centering
\begin{tabular}{clcrc}
\toprule
Nucleus & Sp. & Orbital & \(E^{\mathrm{obs}}\) & Set \\
\midrule

\(^{100}\mathrm{Sn}\) & \(n\) & \(2p_{1/2}\) & -18.38 & E \\
 &  & \(1g_{9/2}\) & -17.93 & E \\
 &  & \(2d_{5/2}\) & -11.13 & E \\
 &  & \(1g_{7/2}\) & -10.93 & E \\
 &  & \(3s_{1/2}\) & -9.30 & E \\
 &  & \(2d_{3/2}\) & -9.20 & E \\
 &  & \(1h_{11/2}\) & -8.60 & E \\
 & \(p\) & \(1f_{5/2}\) & -8.71 & E \\
 &  & \(2p_{3/2}\) & -6.38 & E \\
 &  & \(2p_{1/2}\) & -3.53 & E \\
 &  & \(1g_{9/2}\) & -2.92 & E \\
\midrule

\(^{132}\mathrm{Sn}\) & \(n\) & \(1g_{7/2}\) & -9.75 & ID \\
 &  & \(2d_{5/2}\) & -8.97 & ID \\
 &  & \(3s_{1/2}\) & -7.64 & ID \\
 &  & \(1h_{11/2}\) & -7.54 & E \\
 &  & \(2d_{3/2}\) & -7.31 & ID \\
 &  & \(2f_{7/2}\) & -2.47 & ID \\
 &  & \(3p_{3/2}\) & -1.57 & E \\
 &  & \(1h_{9/2}\) & -0.86 & E \\
 &  & \(2f_{5/2}\) & -0.42 & ID \\
 & \(p\) & \(2p_{1/2}\) & -16.01 & ID \\
 &  & \(1g_{9/2}\) & -15.71 & ID \\
 &  & \(1g_{7/2}\) & -9.68 & ID \\
 &  & \(2d_{5/2}\) & -8.72 & ID \\
 &  & \(2d_{3/2}\) & -6.97 & ID \\
 &  & \(1h_{11/2}\) & -6.89 & ID \\
\midrule

\(^{208}\mathrm{Pb}\) & \(n\) & \(1h_{9/2}\) & -11.40 & ID \\
 &  & \(2f_{7/2}\) & -9.81 & ID \\
 &  & \(1i_{13/2}\) & -9.24 & E \\
 &  & \(3p_{3/2}\) & -8.26 & E \\
 &  & \(2f_{5/2}\) & -7.94 & ID \\
 &  & \(3p_{1/2}\) & -7.37 & E \\
 &  & \(2g_{9/2}\) & -3.94 & E \\
 &  & \(1i_{11/2}\) & -3.16 & E \\
 &  & \(1j_{15/2}\) & -2.51 & ID \\
 &  & \(3d_{5/2}\) & -2.37 & ID \\
 &  & \(4s_{1/2}\) & -1.90 & E \\
 &  & \(2g_{7/2}\) & -1.44 & E \\
 &  & \(3d_{3/2}\) & -1.40 & ID \\
 & \(p\) & \(1g_{7/2}\) & -12.00 & ID \\
 &  & \(2d_{5/2}\) & -9.82 & ID \\
 &  & \(1h_{11/2}\) & -9.36 & E \\
 &  & \(2d_{3/2}\) & -8.36 & ID \\
 &  & \(3s_{1/2}\) & -8.01 & ID \\
 &  & \(1h_{9/2}\) & -3.80 & E \\
 &  & \(2f_{7/2}\) & -2.90 & E \\
 &  & \(1i_{13/2}\) & -2.10 & E \\
 &  & \(2f_{5/2}\) & -0.97 & E \\
 &  & \(3p_{3/2}\) & -0.68 & ID \\
 &  & \(3p_{1/2}\) & -0.16 & ID \\

\bottomrule
\end{tabular}
\end{minipage}

\vspace{0.4em}

\parbox{0.96\textwidth}{
\footnotesize
\textit{Note:}
\(\mathrm{ID}\) denotes one of the \(42\) levels used for parameter
identification, while \(\mathrm{E}\) denotes an evaluation-only level.
}

\end{table*}

\section{Computational Cost}
\label{app:computational_cost}

The computational cost of the physics-informed optimization was measured
during training on a Google Colab instance equipped with an NVIDIA L4
GPU. The experimental PINN was trained for \(17\,000\) epochs,
corresponding to an estimated total training time of
\(9.1~\mathrm{h}\).
Using the configuration in Table~\ref{tab:hpo_search_space}, the
dominant cost arises from evaluating WaveNet and its spatial derivatives
for all selected states and collocation points. For network depth \(L\),
width \(W\), a batch of \(N_{\mathrm{state}}\) states, and
\(N_{\mathrm{coll}}\) sampled points, the leading training cost scales
approximately as
\[
\mathcal{O}\!\left(
N_{\mathrm{epoch}}
N_{\mathrm{state}}
N_{\mathrm{coll}}
LW^{2}
\right).
\]

\section{Bayesian Hyperparameter Optimization}
\label{app:bayesian_hpo}

The final architecture and training settings were selected through
Bayesian hyperparameter optimization \cite{wu2019hyperparameter}. The
search explored network dimensions, learning-rate schedules, quadrature
resolutions, and loss weights.

Each candidate configuration was evaluated using spectral accuracy,
physics-residual convergence, parameter stability, and numerical
robustness. Unstable or nonphysical trials were discarded. The selected
configuration was then fixed for all reported Seminole and Wahlborn
experiments. The explored values and final choices are summarized in
Table~\ref{tab:hpo_search_space}.

\begin{table}[H]
\centering
\caption{
Representative hyperparameter values examined during Bayesian
optimization.
}
\label{tab:hpo_search_space}
\small
\setlength{\tabcolsep}{5pt}
\renewcommand{\arraystretch}{1.15}
\begin{tabular}{lll}
\toprule
Hyperparameter & Values examined & Optimized value\\
\midrule
WaveNet depth
& \(3,4,5\)
& \(5\)\\

WaveNet width
& \(128,256,512\)
& \(256\)\\

ParamNet depth
& \(3,4,5\)
& \(3\)\\

ParamNet width
& \(128,256,512\)
& \(256\)\\

Radial probe points \(N_p\)
& \(64,96,128\)
& \(96\)\\

Decay factor \(\beta\)
& \(0.5, 0.6, 0.7\)
& \(0.6\)\\

WaveNet learning rate
& \(10^{-4},\,5\times10^{-4},\,10^{-3}\)
& \(5\times10^{-4}\)\\

ParamNet learning rate
& \(10^{-4},\,5\times10^{-4},\,10^{-3}\)
& \(10^{-3}\)\\

Radial quadrature \(N_r\)
& \(512,1024,2048\)
& \(1024\)\\

Polar quadrature \(N_\theta\)
& \(256,512,1024\)
& \(512\)\\

Azimuthal quadrature \(N_\phi\)
& \(128,256,512\)
& \(256\)\\

Energy weight \(w_E\)
& \(0.1,0.5,1,5\)
& \(0.5\)\\

Radial-residual weight \(w_R\)
& \(5,7,10\)
& \(10\)\\

Angular-residual weights \(w_\theta\), \(w_\phi\)
& \(3,5,10\)
& \(5\)\\

Boundary weight \(w_{\mathrm{BC}}\)
& \(1,5,10\)
& \(5\)\\

Orthogonality weight \(w_{\mathrm{ORTH}}\)
& \(0,1,5,10\)
& \(5\)\\

Spin--orbit weight \(w_{\mathrm{SO}}\)
& \(0.2,0.5,1\)
& \(0.2\)\\

KL weight \(w_{\mathrm{KL}}\)
& \(10^{-4},10^{-3},10^{-2}\)
& \(10^{-4}\)\\

Latent prior standard deviation
& \(1,2,3\)
& \(2\)\\

WaveNet scheduler factor
& \(0.5,0.6,0.8\)
& \(0.6\)\\

ParamNet scheduler factor
& \(0.5,0.6,0.8\)
& \(0.5\)\\
\bottomrule
\end{tabular}
\end{table}

The final configuration provided the best compromise between spectral
accuracy, physical-residual convergence, parameter stability, and
computational cost. Increasing the network widths or quadrature
resolutions beyond the selected values increased memory consumption and
training time without producing a consistent improvement in the
validation criterion. Smaller networks and coarser quadrature grids
converged more rapidly but produced less stable wavefunction residuals
and parameter estimates.


\FloatBarrier

\bibliographystyle{elsarticle-num-names}
\bibliography{references}

\end{document}